\documentclass[twocolumn,showpacs,prd,aps,tightenlines,floatfix,final]{revtex4}

\pdfoutput=1

\usepackage{graphics,graphicx}
  \DeclareGraphicsExtensions{.pdf}   
  \graphicspath{{final_figures_pdf/}}

\usepackage{color}
\usepackage{amsmath,amssymb}

\usepackage{array}
\newcolumntype{x}[1]{%
>{\centering\hspace{0pt}}p{#1}}%
\newcommand{\tn}{\tabularnewline}

\usepackage{warpcol}
\newcolumntype{L}[1]{>{\pcolbegin{r}{#1}}l<{\pcolend}}
\newcolumntype{R}[1]{>{\pcolbegin{r}{#1}}r<{\pcolend}}

\def \dd{\mathrm{d}}
\newcommand{\parder}[2]{\frac{\partial #1}{\partial #2}}
\newcommand{\bs}[1]{\boldsymbol{#1}}

\begin{document}

\title{
Numerical black hole initial data with low eccentricity \\
based on post-Newtonian orbital parameters
}

\author{Benny~Walther, Bernd~Br\"{u}gmann, Doreen~M\"{u}ller}

\affiliation{Theoretical Physics Institute, 
University of Jena, 07743 Jena, Germany}

\date{May 5, 2009}

\begin{abstract}
Black hole binaries on non-eccentric orbits form an important subclass
of gravitational wave sources, but it is a non-trivial issue to
construct numerical initial data with minimal initial eccentricity for
numerical simulations. We compute post-Newtonian orbital parameters
for quasi-spherical orbits using the method of Buonanno, Chen and
Damour (2006) and examine the resulting eccentricity in numerical
simulations. Four different methods are studied resulting from the
choice of Taylor-expanded or effective-one-body Hamiltonians, and from
two choices for the energy flux. For equal-mass, non-spinning binaries
the approach succeeds in obtaining low eccentricity numerical initial
data with an eccentricity of about $e=0.002$ for rather small initial
separations of $D\gtrsim10M$.  The eccentricity
increases for unequal masses and for spinning black holes, but remains
smaller than that obtained from previous post-Newtonian
approaches. The effective-one-body Hamiltonian offers advantages for
decreasing initial separation as expected, but in the context of this
study also performs significantly better than the Taylor-expanded
Hamiltonian for binaries with spin. For mass ratio 4:1 and vanishing
spin, the eccentricity reaches $e=0.004$. For mass ratio 1:1 and
aligned spins of size $0.85M^2$ the eccentricity is about $e=0.07$ for
the Taylor method and $e=0.014$ for the effective-one-body method.
\end{abstract}

\pacs{
04.25.D-, % Numerical relativity 
04.25.dg, % Numerical studies of black holes and black-hole binaries
04.25.Nx  % Post-Newtonian approximation
}

\maketitle

%%%%%%%%%%%%%%%%%%%%%%%%%%%%%%%%%%%%%%%%%%%%%%%%%%%%%%%%%%%%%%%%%%%%%%%%%

\section{Introduction}
\label{sec:Introduction}

Gravitational wave astronomy holds the promise to open up an entirely
new window into the universe, and a number of ground-based
gravitational wave detectors are now collecting data in the quest for
first direct detection~\cite{Lue06,Abb07a,TatAlt07,AceAlt07}.
Binary black hole (BBH) mergers are expected to be primary sources of
gravitational waves. Theoretical predictions for waves from BBH
mergers are now becoming available for various merger scenarios based
on numerical simulations as well as post-Newtonian (PN) approximation
methods, with increasing quality and range of validity (e.g.\
\cite{Pre07,Bla02}).
Among the likely astrophysical scenarios is that of two black holes
merging at the end-point of a non-eccentric, quasi-circular inspiral.
Although eccentric, non-circular orbits are certainly possible and of
interest as well, non-eccentric orbits are expected to be common since
the emission of gravitational waves reduces eccentricity on time
scales that are short compared to the life time of the binary.

In order to predict waveforms for non-eccentric inspirals,
numerical simulations for the Einstein equations have to face the
issue of how to construct black hole initial data with minimal initial
eccentricity. 
This is a non-trivial task for several reasons. The main reason is
that the initial data have to satisfy the constraint equations of
general relativity. Typical initial data constructions allow us to
specify the ``bare'' parameters of a multiple black hole system
(masses, positions, momenta, and spins), and the field content (the
metric and extrinsic curvature on a hypersurface) is generated by
solving elliptic equations. The solution process changes or dresses
the bare parameters, e.g.\ the physical masses and the proper distance
between the black holes are obtained as part of the solution. There
are therefore two levels of indirectness in the initial data
construction. Discrete parameters are translated into fields, and only
after solving the constraints the physical parameters are known. The
goal is to obtain an accurate approximation to the physical situation
at some given instant of time, which as a matter of principle can only
be an approximation and cannot be exact because the evolution prior to this
time is not available.

With regard to eccentricity, the question is how initial parameters
can be found that result in numerical binary inspirals with minimal
eccentricity. In fact, most numerical simulations have to contend with
a small but sometimes non-negligible amount of eccentricity, and it
becomes a quantitative question to what extent the computation of
faithful gravitational wave templates is hampered by unwanted
eccentricity. 
For example, the Numerical Injection and Analysis (NINJA) project
\cite{AylBakBog09} could benefit from and could evaluate the
importance of low-eccentricity initial configurations.

Several methods are available to find parameters for approximately
circular inspirals. A first approximation is to impose a type of
quasi-circularity or quasi-equilibrium condition on the two black
holes when solving the constraints, which typically is implemented as
an iteration of the orbital parameters and by repeatedly solving the
constraints until the quasi-equilibrium condition is satisfied within
some given error (see e.g.\ \cite{Coo00}).  However, the final judge
of the eccentricity contained in initial data is to perform numerical
evolutions and to measure the resulting eccentricity. This leads to
the immediate suggestion to iterate initial parameters based on the
eccentricity observed in actual evolutions --- which, of course,
became an option only after numerical simulations of several orbits
became possible.  Ref.\
\cite{PfeAlt07} suggests and applies a parameter iteration method based on
evolutions with a pseudospectral code. The eccentricity is estimated
after one or two orbits, and the run is restarted with improved
parameters. The appeal of this method is its generality. It still has
to be determined how well it can work with less accurate finite
differencing codes. In particular, due to artificial waves and gauge
issues it is not clear how many orbits are required to measure the
eccentricity with sufficient accuracy in the numerical simulation. For
short simulations of only a few orbits, several iterations of the
method may be computationally expensive.

Another natural idea in this context is to obtain orbital parameters
from PN methods. Once PN orbital parameters for minimal eccentricity
within the PN approximation are found, they can be used as input for a
constraint solution scheme within the full theory.  Even simple
implementations of this method can work surprisingly well.  For
example in \cite{BruGonHan06} (see
also~\cite{BakCamLou02,BerIyeWil06}), it turns out that already a
simple 3PN formula to compute the tangential momentum $P_t$ for a PN
circular orbit with vanishing radial momentum $P_R$, leads to
approximately circular inspiral initial data when combined with the
puncture method~\cite{BraBru97} to construct binary black hole initial
data.

Typical numerical simulations start shortly before the merger, say at
an initial separation that allows 5 to 15 orbits before merger. In
this regime, the PN approach becomes increasingly inaccurate with
decreasing separation as the dynamic, strong field effects of the
inspiral become severe. In order to improve on the PN parameters of
\cite{BruGonHan06}, we can look to more sophisticated PN approximations, say
involving the effective-one-body (EOB)
approach~\cite{BuoDam98,BuoDam00,DamJarSch00b,Dam01}. 
In particular, the PN method should also supply a
non-vanishing radial momentum $P_R$. Ironically, it is the use of
``quasi-circular'' initial parameters with vanishing radial momentum 
that leads to non-circular, eccentric inspirals, since by definition a
spiral requires a non-vanishing radial momentum.

We note in passing that there also exists a method to translate PN
data including not only parameters $P_t$ and $P_R$, but also the PN
field content to numerical initial data~\cite{TicBruCam02}. The result
is black hole puncture data with more realistic gravitational wave
content~\cite{KelTicCam07}, a feature shared with the periodic
standing wave approximation, e.g.~\cite{HerPri08}. This feature is not
essential for numerical evolutions, but in any case is not implemented
by the methods discussed here.

One way to obtain improved PN parameters for inspirals with small
eccentricity is to start with a simple initial guess
and to perform evolutions of a few hundred
orbits in the PN Hamiltonian formulation until the orbit has
circularized to a high degree due to the emission of gravitational waves
\cite{HusHanGon07}.
In terms of computational effort, note that once
the PN formulas for the Hamiltonian system have been coded, the time
integration is easily handled by a standard Mathematica function in
less than a minute. The elegance of the PN evolution method is that
for any reasonable initial guess the PN evolution leads to a
circularized inspiralling orbit due to the physical process of wave
emission, and this is achieved with little additional effort.

For numerical simulations of equal mass binaries without spin, the
residual eccentricity is reduced by up to a factor of five by the PN
evolution method~\cite{HusHanGon07} compared to simple 2PN
quasi-circular parameters \cite{Kid95,BruGonHan07}, with an
eccentricity $e < 0.002$ at an initial separation of $D=11M$.
For comparison, several equal mass data sets are mentioned in
\cite{HanHusBak09}, with $e=0.0016$ the lowest value used for finite
differencing codes, $e=0.008$ a typical value prior to the recent
improvements, but $e < 5\times10^{-5}$ for the iteration method in a
pseudospectral code~\cite{PfeAlt07}.
For more general binaries with spin and unequal masses, there is some
improvement of a typical initial guess using the PN evolution method,
but the numerical eccentricity is not always as small as one may
wish~\cite{DamNagHan08,CamLouNak08}.  As suggested in~\cite{HusHan08},
it may be possible to add one iteration step to improve the situation
for binaries with spin.

The PN evolution method also demonstrates how large the eccentricity
can be within the PN method itself if certain simple prescriptions for
quasi-circular orbit parameters are used. On the other hand, the
question arises how well a more sophisticated PN method can achieve
orbital parameters for non-eccentric inspirals. In
particular, we want to ask the question whether there is a PN method
that produces orbital parameters directly (without resorting to PN
evolutions) that leads to reduced eccentricity for numerical
simulations. As explained above, we do not look for a method that
reduces the eccentricity to zero, but any PN parameter proposal can be
called successful e.g.\ if it reduces the eccentricity below the level of
changes introduced by the constraint solving process. 
In any case, PN parameters can provide a good initial guess for an 
iteration based on full numerical simulations. 

One of the problems associated with the application of PN methods is
that there is not one, but a multitude of approaches that differ in
strategy as well as in small but important details. For example, two
sets of PN initial parameters may be equivalent up to a certain PN
order, but show differences in numerical simulations, which may be
sensitive to the higher order terms that have been neglected. 

In this paper we focus on the proposal of Buonanno, Chen, and
Damour~\cite{BuoCheDam05} to compute quasi-spherical PN initial data.
This work was based on the Hamiltonian formulation plus the known energy-loss
rate and considered an adiabatic sequence of spherical orbits. That
means the tangential momentum is determined by demanding a constant
binary separation when using the conservative part only. In a second
step the radial component is derived by asking for the rate at which
these spherical orbits are (adiabatically) traversed by including the
energy flux. The approach is capable of incorporating spin-orbit
interactions to leading order, which may present a simple way of
improving initial data for spinning binaries.

The present paper implements the suggestions and the algorithm of
\cite{BuoCheDam05} and modifies it to be more directly applicable to
dynamical variables instead of kinematical ones, i.e.\ ADMTT momenta
instead of velocities. Following \cite{BuoCheDam05} we implement two
variants of PN Hamiltonians, the usual Taylor-expanded and the
EOB version. We also consider two versions of the
energy flux, the classical Taylor-expanded flux for circularly
orbiting bodies and a particular Pad\'{e} approximant of it. When
combined with the two choices of Hamiltonians this results in four
different versions to study.

In Section \ref{sec:pneom} we provide a detailed survey of the
required equations. We further examine the transition from EOB
coordinates to ADMTT-type coordinates, which becomes necessary when
extracting initial data from the EOB Hamiltonian. The initial data
algorithm itself is described in Section \ref{sec:idalgorithm}. Section
\ref{sec:numresults} is dedicated to the exploration of these data in
numerical simulations with the BAM code. By measuring the orbital
eccentricity obtained in numerical simulations we can argue which of
the four data types is most appropriate. The investigation includes
mass ratios 1:1, 2:1, and 4:1 for vanishing spin, several cases of
spin aligned to the orbital angular momentum for mass ratio 1:1, one 
anti-aligned case and one situation with arbitrarily chosen spins.
Especially for binaries with spin the resulting eccentries are found
to be quite large, but initial data involving the
EOB Hamiltonian still give reasonably small eccentricity.

%%%%%%%%%%%%%%%%%%%%%%%%%%%%%%%%%%%%%%%%%%%%%%%%%%%%%%%%%%%%%%%%%%%%%%%%%

\section{Post-Newtonian Equations of motion for spinning binary systems}
\label{sec:pneom}

This section provides the required equations of motion
of a black-hole binary system consisting of objects of masses $m_a$,
positions $\bs{X}_a$, momenta $\bs{P}_a$ and spins $\bs{S}_a$
($a=1,2$). For our purposes it is sufficient to restrict
considerations to the center-of-mass dynamics, where
$\bs{P}\equiv\bs{P}_1=-\bs{P}_2$. Following \cite{BuoCheDam05} we use
a Hamiltonian description in either ADMTT gauge or in EOB coordinates.

The Hamiltonian formalism turns out to be very useful when working
with the canonically conjugate position and momentum variables and in
distinguishing conservative and radiative effects. The system's
equations of motion take the form
\begin{align}
 \frac{\dd\bs{X}}{\dd t}&=\left\{\bs{X},H \right\}=\parder{H}{\bs{P}},\label{eq:nonconservx}\\
 \frac{\dd\bs{P}}{\dd t}&=\left\{\bs{P},H \right\}+\bs{F}=-\parder{H}{\bs{X}}+\bs{F},\label{eq:nonconservp}\\
 \frac{\dd\bs{S}_a}{\dd t}&=\left\{\bs{S}_a,H \right\}=\parder{H}{\bs{S}_a}\times\bs{S}_a,\label{eq:spineom}
\end{align}
where $\bs{F}$ labels the non-conservative force and $\times$
denotes the usual vector cross product. The conservative part consists
of an orbital and a spin contribution and we will use the total
Hamiltonian formed by the sum of three components,
\begin{align}
 H(\bs{X},\bs{P}, \bs{S}_1, \bs{S}_2)=&H^0(\bs{X},\bs{P})+H_\text{SO}(\bs{X},\bs{P},\bs{S}_1,\bs{S}_2)\nonumber\\[6pt]
&+H_\text{SS}(\bs{X},\bs{P},\bs{S}_1,\bs{S}_2).\label{eq:sumhamiltonian}
\end{align}
The individual terms are presented in the subsequent paragraphs.
\vspace*{2.0cm}

\subsection{Orbital Hamiltonian}

We consider two versions of the orbital contribution $H^0$, the
standard 3PN accurate Taylor-expanded Hamiltonian (TH) derived by
Damour, Jaranowski, and
Sch\"{a}fer~\cite{DamJarSch99,DamJarSch00b,DamJarSch00c,DamJarSch01},
and the effective-one-body Hamiltonian (EH) given by Buonanno, Damour,
Jaranowski, and Sch\"{a}fer at 3PN order
in~\cite{BuoDam00,DamJarSch00b} (given first at 2PN
in~\cite{BuoDam98}).
It is crucial to note that the two Hamiltonians are defined for different
coordinate systems. The Taylor-expanded Hamiltonian uses ADMTT-type
coordinates denoted by $\bs{X}, \bs{P}$, where, however, a small deviation from
the actual ADMTT gauge arises at the 3PN-level \cite{DamJarSch99}. 
The EOB coordinates, indicated by
primes $\bs{X}', \bs{P}'$, are related to the ADMTT coordinates by a
canonical transformation. The character of this transformation will be
investigated in Section \ref{sec:ADM2EOB} below. Some equations are
valid for both approaches, TH and EH, and in these cases the primes
are omitted as in
Eqns.\ (\ref{eq:nonconservx})-(\ref{eq:sumhamiltonian}).\\

Both versions of the Hamiltonian will be given explicitly. By
introducing reduced coordinates
$\bs{q}=\bs{X}/(GM)=(\bs{X}_1-\bs{X}_2)/(GM)$ and $\bs{p}=\bs{P}/\mu$,
where $M=m_1+m_2$ and $\mu=m_1m_2/M$, the Taylor-expanded version
reads
\begin{widetext}
\begin{equation}
 H^0_\text{Taylor}(\bs{q},\bs{p})=Mc^2+\mu \left[\hat{H}_{\text{N}}(\bs{q},\bs{p})
 +\frac{1}{c^2}\hat{H}_{1\text{PN}}(\bs{q},\bs{p})+\frac{1}{c^4}\hat{H}_{2\text{PN}}(\bs{q},\bs{p})
+\frac{1}{c^6}\hat{H}_{3\text{PN}}(\bs{q},\bs{p})   \right],\label{eq:pnadmhamiltonian}
\end{equation}
with
\begin{align}
 \hat{H}_{\text{N}}(\bs{q},\bs{p})=&\frac{\bs{p}^2}{2}-\frac{1}{q},\\
 \hat{H}_{1\text{PN}}(\bs{q},\bs{p})=&\frac{1}{8} (3 \nu -1) (\bs{p}^2)^2-\frac12\bigg[ (3+\nu )
   \bs{p}^2+\nu(\bs{n}\bs{p})^2\bigg]\frac{1}{q}+\frac{1}{2 q^2},\label{eq:1pn}
\end{align}
\begin{align}
\hat{H}_{2\text{PN}}(\bs{q},\bs{p})=&\frac{1}{16} \left(1-5 \nu +5 \nu^2\right) (\bs{p}^2)^3 + \frac18\bigg[\left(5-20\nu-3\nu^2\right) (\bs{p}^2)^2-2 \nu ^2 (\bs{n}\bs{p})^2 \bs{p}^2 - 3 \nu ^2 (\bs{n}\bs{p})^4\bigg]\frac{1}{q}\nonumber\\
&+\frac12 \bigg[3 \nu  (\bs{n}\bs{p})^2+(5+8 \nu ) \bs{p}^2\bigg]\frac{1}{q^2} -\frac14\left(1+3\nu\right)\frac{1}{q^3},\\
\hat{H}_{3\text{PN}}(\bs{q},\bs{p})=&\frac{1}{128}\left(-5+35\nu-70\nu^2+35\nu^3\right)(\bs{p}^2)^4+\frac{1}{16}\bigg[(-7+42\nu-53\nu^2-5\nu^3)(\bs{p}^2)^3\nonumber\\
&+(2-3\nu)\nu^2(\bs{n}\bs{p})^2(\bs{p}^2)^2+3(1-\nu)\nu^2(\bs{n}\bs{p})^4\bs{p}^2-5\nu^3(\bs{n}\bs{p})^6\bigg]\frac{1}{q}\nonumber\\
&+\bigg[\frac{1}{16}(-27+136\nu+109\nu^2)(\bs{p}^2)^2+\frac{1}{16}(17+30\nu)\nu(\bs{n}\bs{p})^2\bs{p}^2+\frac{1}{12}(5+43\nu)\nu(\bs{n}\bs{p})^4\bigg]\frac{1}{q^2}\\ &+\bigg\lbrace\bigg[-\frac{25}{8}+\left(\frac{1}{64}\pi^2-\frac{335}{48}\right)\nu-\frac{23}{8}\nu^2 \bigg]\bs{p}^2+\left(-\frac{85}{16}-\frac{3}{64}\pi^2-\frac{7}{4}\nu \right)\nu(\bs{n}\bs{p})^2 \bigg\rbrace\frac{1}{q^3}\nonumber\\ &+\left[\frac18+\left(\frac{109}{12}-\frac{21}{32}\pi^2\right)\nu\right]\frac{1}{q^4},\nonumber
\end{align}\vskip0.2cm%
\noindent where additionally $q=|\bs{q}|$, $\bs{n}=\bs{q}/q$ and $\nu=\mu/M$
have been used. We already plugged in the appropriate values of the
ambiguity parameters $\omega_\text{static}$ and
$\omega_\text{kinetic}$.

The EOB Hamiltonian is given by the expression
\begin{equation}
 H^0_\text{EOB}(\bs{q}',\bs{p}')= Mc^2\sqrt{1+2\nu\left( \frac{H_{\text{eff}}(\bs{q}',\bs{p}')-\mu c^2}{\mu c^2}\right)},
\end{equation} 
with
\begin{align}
H_{\text{eff}}(\bs{q}',\bs{p}')=&\mu c^2\sqrt{A(q')\bigg[1+\frac{\bs{p}'^2}{c^2}+\left(\frac{A(q')}{D(q')}-1\right)\frac{(\bs{n}'\cdot\bs{p}')^2}{c^2}+\frac{1}{c^4}\frac{1}{q'^2}\left(z_1\,(\bs{p}'^2)^2+z_2\,\bs{p}'^2(\bs{n}'\cdot\bs{p}')^2+z_3\,(\bs{n}'\cdot\bs{p}')^4\right)\bigg]},\label{eq:eobhamiltonian}
\end{align}
where $A(\bs{q}')$ is the Pad\'{e}-resummed function
\begin{equation}
 A(q')=\frac{q'^3(8-2\nu)+\frac{1}{c^2}q'^2(a_4+8\nu-16)}{q'^3(8-2\nu)+\frac{1}{c^2}q'^2(a_4+4\nu)+\frac{1}{c^4}q'(2a_4+8\nu)+\frac{1}{c^6}4(a_4+\nu^2)}\label{eq:eobapotentialpade}
\end{equation} 
\end{widetext}
and the remaining quantities are
\begin{align}
 a_4&=\left[\left(\frac{94}{3}-\frac{41}{32}\pi^2\right)-\frac{z_1}{\nu}\right]\nu,\\[2pt]
D(q')&=1-\frac{1}{c^4}\frac{6\nu}{q'^2}+\frac{1}{c^6}\left[7\frac{z_1}{\nu}+\frac{z_2}{\nu}+(3\nu-26)\right]\frac{\nu}{q'^3},\label{eq:eobdpotential}\\[2pt]
z_1&=z_2=0 \text{~~~~~~and~~~~~~~} z_3=2(4-3\nu)\nu.
\end{align}
As before we make use of reduced variables in the formulation of the
EOB Hamiltonian.

\subsection{Spin Hamiltonian}

The spin part will be considered to leading order only. These terms
can be separated into spin-orbit and spin-spin interactions and have
been known for a long time, see for
example~\cite{DamSch88a,BuoCheDam05}. We have
\begin{align}
H_{\text{SO}}&=2\frac{G}{c^2}\frac{\bs{S}_\text{eff}\cdot \bs{L}}{R^3}\label{eq:hso},\\
\bs{S}_{\text{eff}}&=\left(1+\frac34 \frac{m_2}{m_1} \right)\bs{S}_1+\left(1+\frac34 \frac{m_1}{m_2}\right)\bs{S}_2,\\[12pt]
H_\text{SS}&=H_{S_1S_1}+H_{S_1S_2}+H_{S_2S_2},\label{eq:hss}\\
H_{S_1S_2}&=\frac{G}{c^2}\frac{1}{R^3}\Big[3\left(\bs{S}_1\cdot \bs{N}\right)\left(\bs{S}_2\cdot \bs{N}\right)-\left(\bs{S}_1\cdot \bs{S}_2\right)\Big],\label{eq:hs1s2}\\
H_{S_1S_1}&=\frac{G}{c^2}\frac{1}{2R^3}\Big[3\left(\bs{S}_1\cdot \bs{N}\right)\left(\bs{S}_1\cdot \bs{N}\right)-\left(\bs{S}_1\cdot \bs{S}_1\right)\Big]\frac{m_2}{m_1},\label{eq:hsisi}\\
H_{S_2S_2}&=1\rightleftharpoons 2.
\end{align}
Here, $\bs{L}=\bs{X}\times\bs{P}$ is the orbital angular momentum and
$\bs{N}$ is the same unit vector as $\bs{n}$ above.  
We assume the Newton-Wigner spin supplementary condition, which provides the
notion of the spin vector and affects the definition of the bodies'
worldlines.

In what follows Eqns.~(\ref{eq:hso})-(\ref{eq:hsisi}) are presumed to
hold for ADMTT as well as EOB coordinates. This, however, is an
approximation as already noted in~\cite{BuoCheDam05}. There exists a
more sophisticated and potentially more accurate way of including spin
effects in the EOB framework~\cite{Dam01}, but the Hamiltonian
associated with this approach contains spin-orbit and spin-spin
effects mixed in a complicated way, and it seems that these terms
cannot be separated in a consistent manner.  The initial-data
algorithm we present, however, relies on neglecting the
spin-spin terms for the sake of obtaining spherical orbits. Hence, in
the EOB case we follow~\cite{BuoCheDam05} by just employing the same
simple, additive spin Hamiltonian as in the ADM framework. We discuss
the consequences of this approximation below when analyzing the
canonical transformation between the two coordinate systems.

\subsection{Radiation-reaction force}

The radiation-reaction-force term $\bs{F}$ to be applied in
Eq.~(\ref{eq:nonconservp}) is the expression derived
by~\cite{BuoCheDam05}. It has the feature that it is
formulated in terms of dynamical quantities ($\bs{X}$, $\bs{P}$)
instead of kinematical ones ($\bs{X}$, $\dot{\bs{X}}$). It also
includes spin interactions to linear order and reads
\begin{align}
\bs{F}=&\frac{1}{\omega|\bs{L}|}\frac{\dd E}{\dd t}\bs{P}+\frac{8}{15}\nu^2\frac{v_{\omega}^8}{\bs{L}^2R}\bigg\lbrace\bigg(61+48\frac{m_2}{m_1}\bigg)\bs{P}\!\cdot\!\bs{S}_1\nonumber\\ &+\bigg(61+48\frac{m_1}{m_2}\bigg)\bs{P}\!\cdot\!\bs{S}_2\bigg\rbrace\bs{L}.\label{eq:reactionforce}
\end{align}
The invariant velocity parameter
\begin{equation}
v_\omega=\left(\frac{GM\omega}{c^3}\right)^{1/3}
\end{equation}
based on the orbital frequency $\omega=\dot{\varphi}$ has been
introduced, although this brings the kinematical quantities back into
play.

Eq.~(\ref{eq:reactionforce}) is valid for (quasi-)circular orbits
only, which is a result of simplifications in its derivation and
due to the fact that the quantity $\dd E/\dd t$ is the
well-known energy-flux function of \emph{circularly} orbiting masses
\cite{BlaSch89,JunSch92,BuoCheDam05,BlaFayIye01,BlaFayIye05,BlaDamEsp04}. In
particular, its standard
form (up to 3.5PN order) does not account for any eccentricity
parameter,
\begin{align}
\frac{\dd E}{\dd t}=&-\frac{32}{5}\nu^2v_{\omega}^{10}\bigg\lbrace 1+f_2(\nu)v_{\omega}^2+[f_3(\nu)+f_{3\text{SO}}]v_{\omega}^3\nonumber\\
&+[f_4(\nu)+f_{4\text{SS}}]v_{\omega}^4+f_5(\nu)v_{\omega}^5+f_6(\nu)v_{\omega}^6\label{eq:taylorflux}\\
&+f_{l6}\,v_{\omega}^6\,\ln(4v_{\omega})+ f_7(\nu)v_{\omega}^7\bigg\rbrace.\nonumber
\end{align}
These equations are also available for non-vanishing eccentricity in
\cite{AruBlaIye07}.
The expansion coefficients read
\begin{equation}\label{eq:edotcoeffs}
\begin{aligned}
f_2(\nu)=&-\frac{1247}{336}-\frac{35}{12}\nu,\\[12pt]
f_3(\nu)=&\,4\pi,\\[12pt]
f_4(\nu)=&-\frac{44\,711}{9072}+\frac{9271}{504}\nu+\frac{65}{18}\nu^2,\\[12pt]
f_5(\nu)=&-\left(\frac{8191}{672}+\frac{583}{24}\nu\right)\pi,\\[12pt]
f_6(\nu)=&\frac{6\,643\,739\,519}{69\,854\,400}+\frac{16}{3}\pi^2-\frac{1712}{105}\gamma_\text{E}\\
&+\left(-\frac{134\,543}{7776}+\frac{41}{48}\pi^2 \right)\nu -\frac{94\,403}{3024}\nu^2-\frac{775}{324}\nu^3,\\[12pt]
f_{l6}=&-\frac{1712}{105},\\[12pt]
f_7(\nu)=&\left(-\frac{16\,285}{504}+\frac{214\,745}{1728}\nu+\frac{193\,385}{3024}\nu^2\right)\pi,\\[12pt]
f_{3\text{SO}}=&-\!\!\left(\frac{11}{4}+\frac{5}{4}\frac{m_2}{m_1}\right)\!\frac{\hat{\bs{L}}\cdot\bs{S}_1}{M^2}-\left(\frac{11}{4}+\frac{5}{4}\frac{m_1}{m_2}\right)\!\frac{\hat{\bs{L}}\cdot\bs{S}_2}{M^2},\\[12pt]
f_{4\text{SS}}=&\frac{\nu}{48\,m_1^2\,m_2^2}\Big[289(\hat{\bs{L}}\cdot\bs{S}_1)(\hat{\bs{L}}\cdot\bs{S}_2)-103(\bs{S}_1\cdot\bs{S}_2)\Big]\\
&+\mathcal{O}(\bs{S}_1^2)+\mathcal{O}(\bs{S}_2^2),
\end{aligned}
\end{equation}
where $\gamma_\text{E}=0.577215\ldots$ is Euler's gamma and
$\hat{\bs{L}}$ gives the unit vector in the direction of
$\bs{L}$. Note that the term $f_{4\text{SS}}$ is not fully known yet,
but we will not be using it anyway.

The force expression (\ref{eq:reactionforce}) actually does not enter
the initial data algorithm presented in Section \ref{sec:idalgorithm}
below, however the energy flux (\ref{eq:taylorflux}) does. The fact
that the flux is unable to describe eccentricity damping properly is
not supposed to affect the algorithm. This is because we will merely
ask for the shrinkage rate of \emph{circular} (resp.\ spherical)
orbits at \emph{one} particular radius. In contrast, the application
of the flux and the force $\bs{F}$ in PN simulations with noticeable
(e.g.\ spin induced) eccentricity might give inaccurate results. In
such cases we may want to consider time dependent flux expressions or
time averages over elliptical orbits \cite{BlaSch89}.

Concerning the PN convergence of the flux function, several authors
have shown that the Pad\'{e}-resummation technique is capable of
improving convergence, in particular in the test-mass limit. A recent
paper, however, claims that no advantages can be found using the
Pad\'{e}-resummation approach for several specific
examples~\cite{MroKidTeu08}.
In order to examine this issue in the context of our eccentricity
study, we implement the standard flux (\ref{eq:taylorflux}), but also
consider a \emph{direct P-approximant} of the flux. The final
expression is too lengthy to be shown here. Since
computer-algebra software provides an efficient way to perform the
resummation, we restrict ourselves to a brief summary of the required
procedure. A more detailed description can be found
in~\cite{DamIyeSat97,PorSat05}.

Pad\'{e}'s method defines the resummation of a plain Taylor
series. Referring to (\ref{eq:taylorflux}), we first separate the
spin-dependent expressions and only resum the non-spinning part. In
the remainder we find a logarithmic term which is factored out in a
normalized form before the actual resummation (Eq.\ (38) in
\cite{PorSat05}). The normalization parameter $x_\text{LSO}$ can be
taken from \cite{DamIyeSat97}, Eq.~(3.23). Another
factorization (Eqns.\ (39) and (44) in \cite{PorSat05}) takes care of
a putative pole in the expansion near the light ring, whose position
$v_{\omega\text{~light ring}}$ is given by Eq.\ (3.22) in
\cite{DamIyeSat97} to 2PN accuracy. Thus, finally, one factor of the
flux function is a plain Taylor expansion and is resummed with a
near-diagonal (lower-diagonal) Pad\'{e} approximation.

As a result we have a choice between two versions of $\dd E/\dd t$
that are to be inserted in Eq.~(\ref{eq:reactionforce}). We will refer
to the different fluxes as Taylor flux (TF) and Pad\'{e} flux (PF).

\subsection{\label{sec:ADM2EOB}Mapping EOB to ADMTT coordinates}

The link between ADM and EOB coordinates is given by a canonical transformation 
with generating function $g(\bs{q},\bs{q}')$,
\begin{equation}
 p_i\,\dd q^i=p'_i\,\dd q'^i+\dd g(\bs{q},\bs{q}'),
\end{equation}
where $g_\text{id}(\bs{q},\bs{q}')=0$ would correspond to the identity
transformation \cite{BuoDam98,DamJarSch00b}. Via a Legendre
transformation we arrive at a more convenient generator
$\tilde{G}(\bs{q},\bs{p}')=g(\bs{q},\bs{q}')+p'_i\,q'^i$ yielding
\begin{equation}
 p_i\,\dd q^i+q'^i\,\dd p'_i=\dd\tilde{G}(\bs{q},\bs{p}').
\end{equation}
Decomposition into the identity map plus PN corrections,
\begin{align}
\tilde{G}(\bs{q},\bs{p}')=&\,q^ip'_i+G(\bs{q},\bs{p}'),\\
G(\bs{q},\bs{p}')=&\,\frac{1}{c^2}G_\text{1PN}(\bs{q},\bs{p}')+\frac{1}{c^4}G_\text{2PN}(\bs{q},\bs{p}')\nonumber\\
&+\,\frac{1}{c^6}G_\text{3PN}(\bs{q},\bs{p}')\,,\label{eq:gexpansionpn}
\end{align}
was the key to the determination of the generating function, which was 
accomplished by \cite{BuoDam98,DamJarSch00b}. Their result, which is not unique
when including the 3PN level, reads
\begin{align}
 G_\text{1PN}(\bs{q},\bs{p}')=&\,(\bs{q}\cdot\bs{p}')\left(a_1\bs{p}'^2+\frac{a_2}{q}\right)\label{eq:a1pn},\\
 G_\text{2PN}(\bs{q},\bs{p}')=&\,(\bs{q}\cdot\bs{p}') \bigg[  b_1\bs{p}'^4+\frac{1}{q} \left( b_2\bs{p}'^2+b_3(\bs{n}\cdot\bs{p}')^2 \right)\nonumber\\
&\,+\frac{b_4}{q^2}\bigg],\label{eq:g2pn}\\
G_\text{3PN}(\bs{q},\bs{p}')=&(\bs{q}\cdot\bs{p}') \bigg[  c_1\bs{p}'^6+\frac{1}{q} \big( c_2\bs{p}'^4+c_3\bs{p}'^2(\bs{n}\cdot\bs{p}')^2\nonumber\\
&\,+\,c_4(\bs{n}\cdot\bs{p}')^4\big)+\frac{1}{q^2}\left(c_5\bs{p}'^2+c_6(\bs{n}\cdot\bs{p}')^2\right)\nonumber\\
&\,+\,\frac{c_7}{q^3}\bigg],\label{eq:g3pn}
\end{align}
with
\begin{align}
a_1&=-\frac{\nu}{2}, & a_2&=1+\frac{\nu}{2},\\[10pt]
b_1&= \frac18(\nu+3\nu^2), & b_2&=\frac18(2\nu-5\nu^2),\nonumber\\
b_3&= \frac18(8\nu+3\nu^2), & b_4&= \frac14(1-7\nu+\nu^2),
\end{align}
\begin{align}
c_1&=-\frac{1}{16}(1+3\nu+5\nu^2)\nu\nonumber,\\
c_2&=-\frac{1}{16}(1+2\nu-11\nu^2)\nu\nonumber,\\
c_3&=-\frac{1}{24}(12+48\nu+23\nu^2)\nu\nonumber,\\
c_4&=\phantom{-}\frac{1}{16}(24+7\nu)\nu^2,\label{eq:gcoeffs}\\
c_5&=-\frac{1}{16}(13-16\nu+6\nu^2)\nu,\nonumber\\
c_6&=-\frac{1}{48}(115+116\nu-26\nu^2)\nu,\nonumber\\
c_7&=-\left(\frac{1}{64}\pi^2+\frac{155}{24}\right)\nu+\frac{3}{8}\nu^2+\frac18\nu^3.\nonumber
\end{align}

The sought-after transformation EOB $\rightarrow$ ADM is
\begin{align}
 q^i&=q'^i-\parder{G(\bs{q},\bs{p}')}{p'_i},& p_i&=p'_i+\parder{G(\bs{q},\bs{p}')}{q^i},\label{eq:canonictraforelationback}
\end{align}
which gives an implicit relation that can be solved
numerically by a root-finding algorithm.  Another possibility to
resolve the implicity is to repeatedly plug in the first equation in
order to eliminate the unknown value $\bs{q}$ to sufficient PN order,
yielding an explicit formula. By rearranging
(\ref{eq:canonictraforelationback}) we can calculate the inverse
transformation, ADM $\rightarrow$ EOB, in an analogous manner.
The inverse transformation is convenient when discussing properties of
the transformation and the associated errors, for example we can start
with ADM data from the class of circularly orbiting BH binaries and
study the result on the EOB side.

In order to illustrate the effect of the transformation, we consider
the equal-mass case and choose, as a result from our initial data
algorithm, the values
\begin{align}
\bs{q}&=(10, 0, 0),\nonumber\\
\bs{p}&=(-0.00394009, 0.384516, 0),
\label{eq:qadmsetting}
\end{align}
for a separation of $D=10M$ (with $M=c=G=1$). This translates to
\begin{align}
\bs{q}'&=(10.9445, 0.00318625, 0),\nonumber\\
\bs{p}'&=(-0.00397814, 0.351514, 0).
\end{align}
when using the explicit version of ADM $\rightarrow$ EOB at 3PN
order. Note that the transformation mixes corresponding components of
$(\bs{q}, \bs{p})$ and $(\bs{q}', \bs{p}')$, so the non-vanishing
$y$-component of $\bs{p}$ causes a non-zero $y$-component of
$\bs{q}'$. The intermixture, however, is not a contradiction to the role 
of $\bs{q}$ and $\bs{q}'$ as position variables and $\bs{p}$ and $\bs{p}'$
as the conjugate momenta.

\begin{figure*}[!ht] % 2-column figure
\centering
\includegraphics[width=8.3cm,viewport=84 576 320
768]{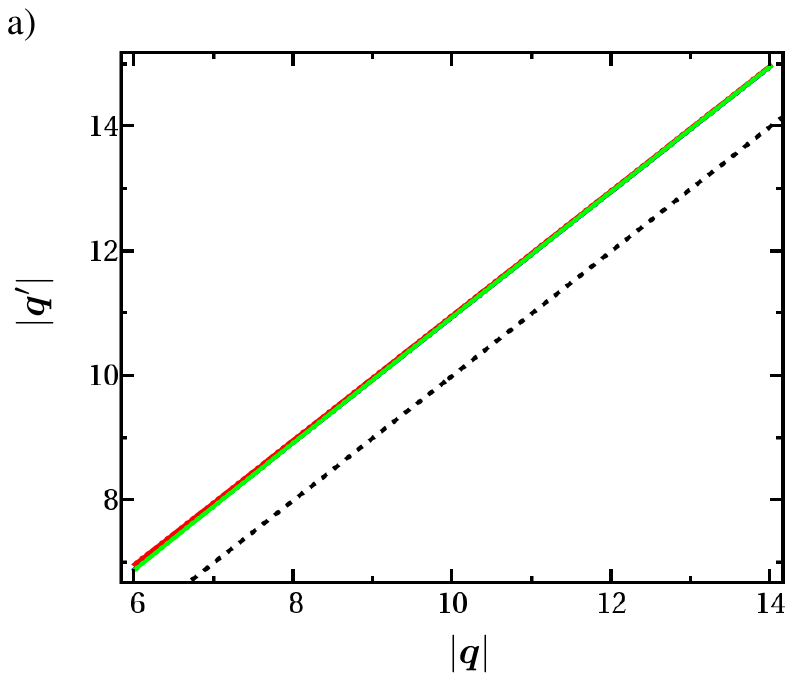}
\hskip0.8cm
\includegraphics[width=8.3cm,viewport=84 576 320
768]{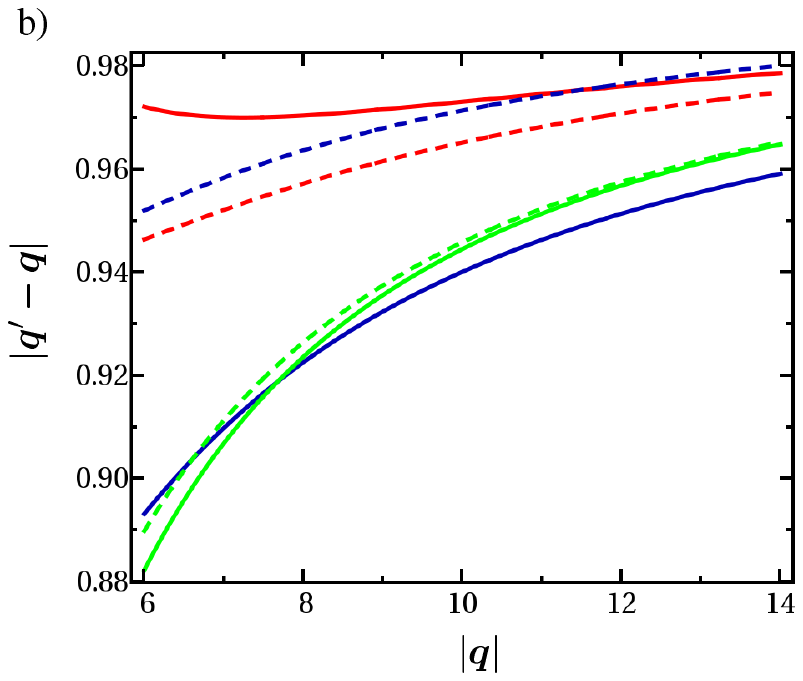}
\\[0.4cm]
\includegraphics[width=8.3cm,viewport=84 576 320
768]{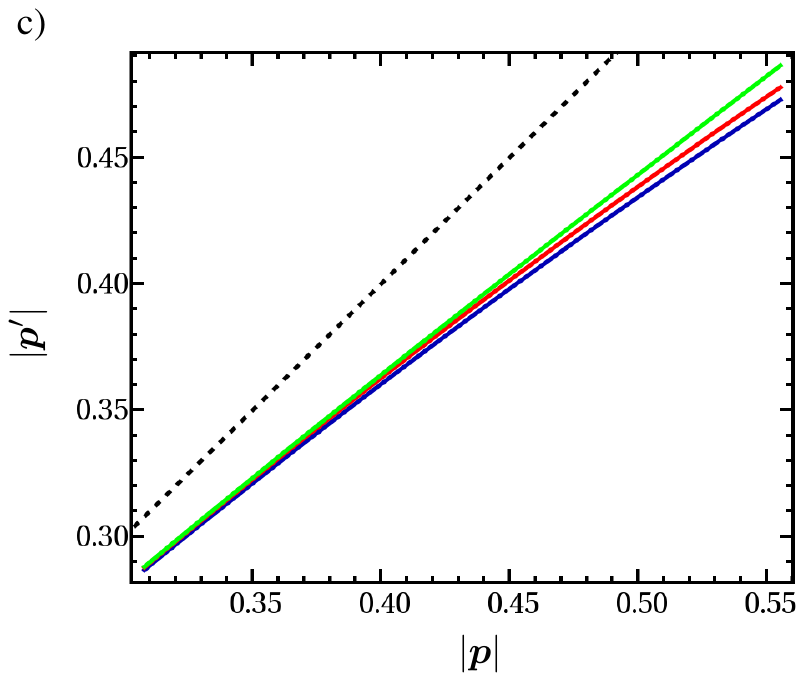}
\hskip0.8cm
\includegraphics[width=8.3cm,viewport=73 370 309
562]{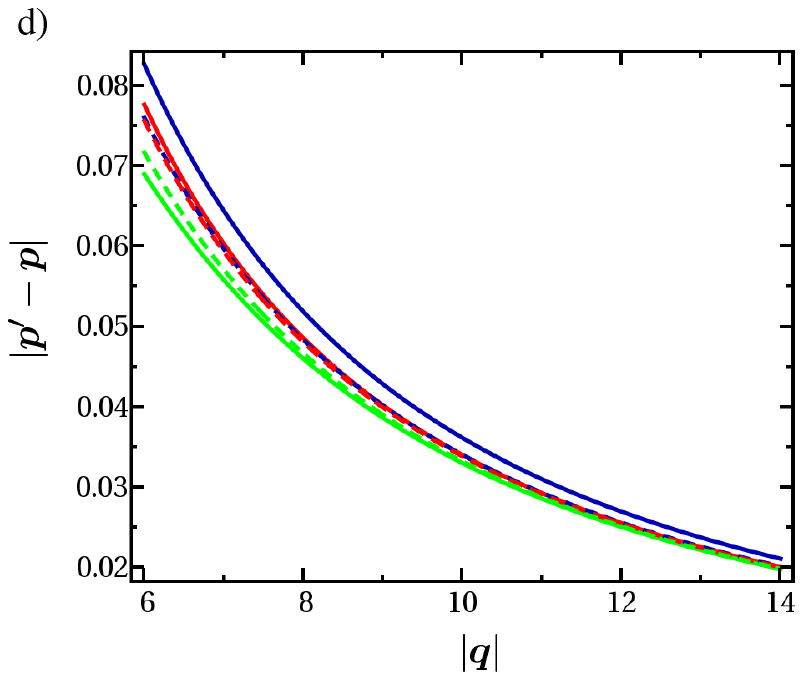}
\\[0.4cm]
\includegraphics[width=9.3cm,viewport=84 724 349
758]{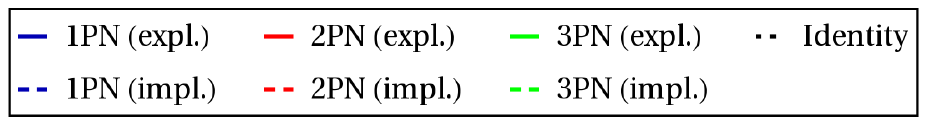}
\caption{
(Color online)
Character of the canonical transformation ADM $\rightarrow$ EOB for
spherical initial data. The dotted line indicates the identity map
$\bs{q}'=\bs{q}$ and $\bs{p}'=\bs{p}$. The plots on the right compare
the explicit and implicit solution method of the
transformation (\ref{eq:canonictraforelationback}) and apparently there is
convergence between the two versions with increasing PN
order. However, convergence is not achieved conclusively, and the
observed differences are used to define an error bar for the
transformation. 
Since the explicit transformation is an approximation (truncated at finite
order) to the implicit relation, it is less accurate.
}
\label{fig:eobvsadmcoordinates}
\end{figure*}

For equal masses, Fig.~\ref{fig:eobvsadmcoordinates} illustrates the
actual relation between both coordinate types in the range from 6$M$ to
14$M$, imposing non-eccentric momenta throughout. In particular,
Fig.~\ref{fig:eobvsadmcoordinates}\,a reveals that $|\bs{q}|$ and
$|\bs{q}'|$ (resp. $R$ and $R'$) lie apart from each other by almost
$1M$.

However, the validity of the canonical transformation depends on the
assumptions $|\bs{q}'-\bs{q}|/|\bs{q}|\ll 1$ and
$|\bs{p}'-\bs{p}|/|\bs{p}|\ll 1$. The fulfillment of these conditions
can be considered critical for the close binary separations under
consideration, which might result in severe problems when using the
spin and flux equations with EOB variables. Namely, as we have just
found there is a significant difference between the radial distances
$R$ and $R'$, as well as some deviation in the direction vectors
$\bs{N}$, $\bs{N}'$. These quantities appear in
Eqns.~(\ref{eq:hso})-(\ref{eq:hsisi}). However, only (\ref{eq:hso}) is
incorporated in the initial data algorithm. Moreover, the
transformation $\dd\bs{p}\wedge\dd\bs{q}=\dd\bs{p}'\wedge\dd\bs{q}'$
corresponds to $\dd H\wedge \dd t=\dd H'\wedge \dd t'$ and implies a
change in the time scale \cite{BuoDam98}. This also affects the
orbital frequency $\omega$ which is used within the flux function
(\ref{eq:taylorflux}). The quantity $\bs{L}$, however, is the same in
both systems. The notion of spin vectors is supposed to be unaltered
up to 1PN level, but modifications occur at higher PN orders. For a
more detailed investigation we refer to the recently
published reference~\cite{DamJarSch08}, where the spin vectors are
made subject of an additional canonical transformation.

Although we are aware of these issues, we note that applying suitable
corrections in order to avoid these inconsistencies would be rather
involved and would probably be unnecessarily sophisticated for our
purpose. As already discussed, further issues arise anyway when
translating the initial parameters obtained this way into the actual
initial data (i.e.\ metric and curvature fields). In the end, the
measured eccentricity decides whether our data -- obtained under 
the specified simplifications -- are appropriate.

\section{Algorithm for quasi-spherical initial data}
\label{sec:idalgorithm}

If we take into account only the conservative part of the orbital
post-Newtonian dynamics, $H^0$, together with the leading-order spin-orbit
interaction term, $H_\text{SO}$, it is possible to find solutions which feature
a constant coordinate separation $R$. These solutions are called spherical
orbits.
In general, this would no longer be true if we additionally included spin-spin
coupling via $H_\text{SS}$, although there might exist very special spin
constellations for which a spherical orbit can be found.
In the following we only consider spin-orbit coupling (with one
exception mentionend below).

Imposing spin-orbit coupling we find two additional conserved
quantities, the magnitude of the angular momentum $L=|\bs{L}|$ and
\begin{equation}
\chi_L=\frac{c}{GM^2}\,\bs{S}_\text{eff}\cdot \hat{\bs{L}},\label{eq:childefinition}
\end{equation}
which gives the projection of the spins onto the direction of $\bs{L}$. Changing
 to polar coordinates with $\bs{L}$ pointing along the $z$-axis, the total
Hamiltonian depends on only four quantities,
\begin{equation}
 H(R,P_R,L,\chi_L)=H^0(R,P_R,L)+2\frac{G}{c^2}\,\frac{GM^2}{c}\,\frac{L\;\chi_L}{R^3}\,,\label{eq:spinorbithamiltonian}
\end{equation}
where we used the relation
\begin{align}
 \bs{P}^2&=P_R^2+P_t^2=P_R^2+\frac{P_\varphi^2}{R^2}
\end{align}
and $P_\varphi=L$. Note that only the $z$-components of the spins
enter. Beyond these considerations for the conservative part,
radiation reaction will be included separately.

The challenge for the initial data algorithm is to find tangential
and radial momentum of the objects for a given $R$ and $\chi_L$ such
that no radial oscillations occur during evolution with the
dissipative PN equations of motion. 

Note that demanding a certain initial separation $R$ is a gauge
dependent statement, which, however, is convenient for our purposes
since the ADMTT gauge resembles the 1+log/gamma-freezing gauge used in
standard puncture evolutions.
Under these circumstances, the algorithm takes a somewhat simpler form
when using $R$ in order to specify the binary separation than 
when specifying the invariant orbital frequency $\omega$ instead, as
originally done by Buonanno et al.\ \cite{BuoCheDam05}. Furthermore,
it eases the specification of spins, which can now be set with
respect to the dynamical quantity
$\bs{\hat{L}}_0=\bs{X}_0\times\bs{P}_0/\vert\bs{X}_0\times\bs{P}_0\vert$
instead of the kinematical variable
$[\bs{\hat{L}}_\text{N}]_0=\bs{X}_0\times\bs{\dot{X}}_0/\vert\bs{X}_0\times\bs{\dot{X}}_0\vert$.
We use the subscript ``0'' to stress that the initial value is meant.

The algorithm itself consists of two steps: (i) Determine $[P_t]_0$
as required for spherical orbits by considering the
conservative motion only. (ii) Calculate the component $[P_R]_0$ by
means of the energy flux in order to get a quasi-spherical inspiral.\\

In the first step of the procedure we determine $[P_t]_0$, or
equivalently $L_0=R_0\cdot [P_t]_0$, by demanding constant radius,
\begin{equation}
0\stackrel{!}{=}[\dot{R}]_0=\left[\parder{H(R,P_R,L,\chi_L)}{P_R}\right]_0.
\end{equation} 
Since $P_R$ enters quadratically into $H$ this is equivalent to
\begin{equation}
[P_R]_0=0.
\end{equation}
The value of $\chi_L$ is known and thus there is only one remaining
degree of freedom, namely the value of $L_0$. It is fixed by
demanding
\begin{equation}
0\stackrel{!}{=}[\dot{P}_R]_0=-\left[\parder{H(R,P_R,L,\chi_L)}{R}\right]_0,\label{eq:prdotequalszerocondition}
\end{equation} 
a condition which has to be solved numerically. A starting value for
an iteration procedure can be provided by the 3PN-accurate tangential
momentum for circular orbits of non-spinning objects,
\begin{align}
P_t^\text{3PN}=&\mu\Bigg[\left(\frac{GM}{D}\right)^{1/2}+\frac{1}{c^2}2\left(\frac{GM}{D}\right)^{3/2}+\frac{1}{c^4}\frac{1}{16}(42\nonumber\\
&-43\nu)\left(\frac{GM}{D}\right)^{5/2}+\frac{1}{c^6}\frac{1}{128}[480+(163\pi^2\nonumber\\
&-4556)\nu+104\nu^2]\left(\frac{GM}{D}\right)^{7/2}\Bigg],\label{eq:hannamptangest}
\end{align} 
a formula given in~\cite{BruGonHan06}.

\begin{table*}[!ht]
\centering
\begin{tabular}{x{2.5cm}x{1.0cm}x{1.0cm}x{1.0cm}x{1.0cm}x{1.0cm}x{1.0cm}x{1.0cm}
x{1.0cm}x{1.0cm}}
\hline
\hline
\multicolumn{1}{c}{\rule{0pt}{11pt}Mass ratio} & \multicolumn{2}{l}{$~R_0~/~M$}
&&&&&&&\tn 
& \multicolumn{1}{c}{6} & \multicolumn{1}{c}{7} & \multicolumn{1}{c}{8} &
\multicolumn{1}{c}{9} & \multicolumn{1}{c}{10} & \multicolumn{1}{c}{11} &
\multicolumn{1}{c}{12} & \multicolumn{1}{c}{13} & \multicolumn{1}{c}{14}\tn
\hline
\rule{0pt}{14pt}\phantom{1}1:1  & 3.1\phantom{0} & 1.8\phantom{0} &
1.1\phantom{0} & 0.78 & 0.55 & 0.40 & 0.30 & 0.23 & 0.18\tn
\phantom{1}4:1  & 2.2\phantom{0} & 1.2\phantom{0} & 0.74 & 0.49 & 0.34 & 0.25 &
0.19 & 0.14 & 0.11\tn
10:1 & 0.99 & 0.55 & 0.34 & 0.21 & 0.15 & 0.11 & 0.08 & 0.06 & 0.05\tn
\hline
\hline
\end{tabular} 
\caption{
Estimate of the relative error in percent for EOB-based initial data
$P_t$ and $P_R$ at various binary separations and mass
ratios.  
}
\label{tab:relativeerrorinmomentum}
\end{table*}%\vskip0.5cm

The second step incorporates radiation reaction and consequently a
non-vanishing $\dot{R}$. Assuming adiabatic inspiral, that is a
sequence of spherical orbits whose shrinkage in radius is determined
by the energy flux, one finds
\begin{equation}
[\dot{R}]_0=\frac{[\dd E/\dd t]_0}{[(\dd E/\dd R)_\text{sph}]_0}\,,\label{eq:rdotinstep2}
\end{equation} 
where $\dd E/\dd t$ is given by Eq.\ (\ref{eq:taylorflux}) or its
Pad\'e resummation. Here the knowledge of the initial orbital
frequency $\omega_0$ is required and can be provided by the evolution
equation
\begin{equation}
\omega_0=[\dot{\varphi}]_0=\left[\parder{H}{P_\varphi}\right]_0=\left[\parder{H}{L}\right]_0.
\end{equation}
The initial conditions used in the RHS of this equation,
i.e.\ the components of the momentum, are taken from the first step,
in particular $[P_R]_0=0$. The other quantity appearing in
(\ref{eq:rdotinstep2}) is the energy difference between neighboring
spherical orbits and can be calculated as
\begin{align}
\left(\frac{\dd E}{\dd R}\right)_\text{\!sph}=&\parder{H}{R}\left(\frac{\dd R}{\dd R}\right)_\text{\!sph}+\parder{H}{P_R}\left(\frac{\dd P_R}{\dd R}\right)_\text{\!sph}\nonumber\\
&+\parder{H}{L}\left(\frac{\dd L}{\dd R}\right)_\text{\!sph}+\parder{H}{\chi_L}\left(\frac{\dd\chi_L}{\dd R}\right)_\text{\!sph}.\label{eq:debydrsph1}
\end{align}
The first two terms vanish, because for spherical orbits $\partial
H/\partial R=-P_R=0$ holds and consequently $\dd P_R=0$, too. The
other two summands are problematic in that there appear the unknowns
$(\dd L/\dd R)_\text{sph}$ and $(\dd\chi_L/\dd R)_\text{sph}$. We are
able to substitute the first of them by considering
\begin{align}
0&=\left(\frac{\dd}{\dd R}\parder{H}{R} \right)_\text{\!sph}\nonumber\\
&=\frac{\partial^2H}{\partial R^2}+\frac{\partial^2H}{\partial L\, \partial R}\left(\frac{\dd L}{\dd R}\right)_\text{\!sph}+\frac{\partial^2H}{\partial \chi_L\, \partial R}\left(\frac{\dd \chi_L}{\dd R}\right)_\text{\!sph},
\end{align}
where we have omitted vanishing terms. Rearranging for $(\dd L/\dd R)_\text{sph}$ and inserting into (\ref{eq:debydrsph1}) we obtain
\begin{align}
\left(\frac{\dd E}{\dd R}\right)_\text{\!sph}=&-\frac{\left(\parder{H}{L}\right)\left(\parder{^2H}{R^2}\right)}{\left(\parder{^2H}{R\,\partial L}\right)}+\Bigg[\left(\parder{H}{\chi_L}\right)\nonumber\\
&-\frac{\left(\parder{H}{L}\right)\left(\parder{^2H}{R\,\partial\chi_L}\right)}{\left(\parder{^2H}{R\,\partial L}\right)} \Bigg]\left(\frac{\dd\chi_L}{\dd R}\right)_\text{\!sph},\nonumber\\[12pt]
\left[\left(\frac{\dd E}{\dd R}\right)_\text{\!sph}\right]_0\approx & -\frac{\left[\left(\parder{H}{L}\right)\right]_0\left[\left(\parder{^2H}{R^2}\right)\right]_0}{\left[\left(\parder{^2H}{R\,\partial L}\right)\right]_0}.\label{eq:debydrsph}
\end{align}
The second part has been neglected as $\chi_L$ is supposed to be an
almost conserved quantity even though radiation reaction is turned
on. According to the analysis of \cite{BuoCheDam05} the corresponding
error is of 3PN order. With $[\dot{R}]_0$ in hand we can exploit the
abovementioned quadratic appearance of $P_R$ within the Hamiltonian to
obtain
\begin{align}
[P_R]_0=\frac{[\dot{R}]_0}{\left[\frac{1}{P_R}\parder{H}{P_R}\right]_{0, P_R\rightarrow 0}}=\frac{[\dot{R}]_0}{2\left[\parder{H}{(P_R^2)}\right]_0}.
\end{align}

Hence, the algorithm gives $[P_t]_0$ and $[P_R]_0$ in a coordinate
system where $\hat{\bs{L}}_0$ (by definition) represents the
$z$-axis. We omit the subscript ``0'' from here on as it is clear that
we refer to initial data. Furthermore we use the labels $R$ and $D$,
denoting the binary separation, interchangeably.

The procedure just described is valid for both types of
Hamiltonians. For the Taylor-expanded version we immediately obtain
the final result in ADMTT gauge. Employing the EOB approach on the
other hand is rather more involved. We shall briefly illustrate why.

First, recall that in the EOB case the above algorithm is applied for
the primed variables, which represents a simplification as we
discussed above. However, in the end we want to obtain the result in
ADMTT coordinates. Moreover, we wish to specify the ADMTT distance,
i.e.\ $\bs{X}$, at which the initial momentum is to be determined. The
canonical transformation ADM $\leftrightarrow$ EOB only mediates
between the whole set of variables
$(\bs{X},\bs{P})\leftrightarrow(\bs{X}',\bs{P}')$, thus providing the
corresponding value of $\bs{X}'$ for use within the algorithm would
already require the knowledge of the sought-after solution
$\bs{P}$. Hence, an additional step becomes necessary, for example a
shooting method, to achieve a match with the given ADM distance. We
take care of this for our actual initial data computations in order to
compare numerical results at identical values of $\bs{X}$.

\begin{figure*}[!ht] % 2-column figure 
\centering
\includegraphics[width=8.3cm,viewport=84 595 320 761]{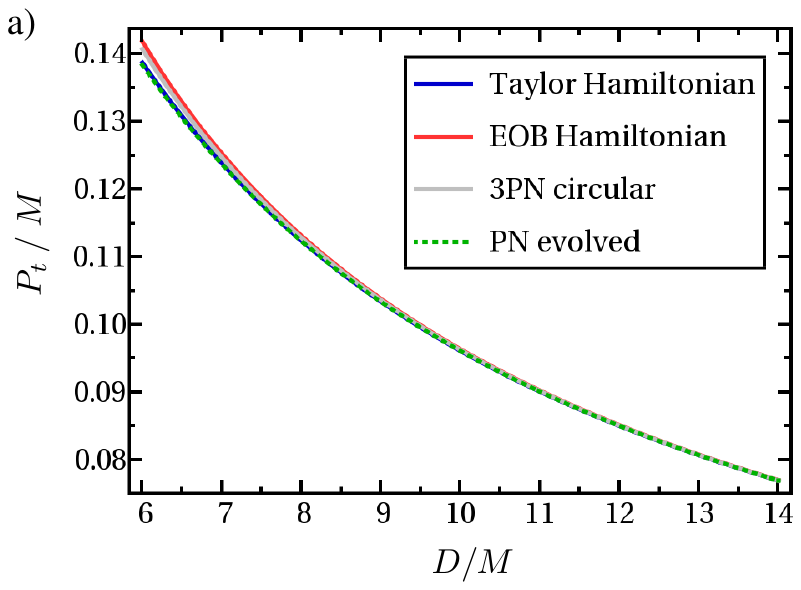}%{pt_em_nospin}
\hskip0.8cm
\includegraphics[width=8.3cm,viewport=81 595 320
761]{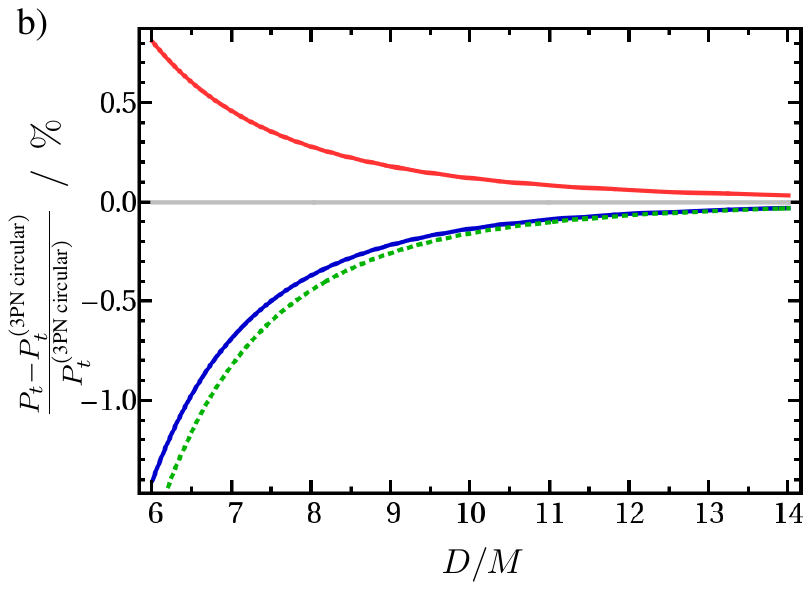}
\caption{(Color online) Tangential momentum obtained with the initial data
algorithm for an 
equal-mass binary ($M=1$, no spins). Also shown are reference functions ``3PN
circular'' from \cite{BruGonHan06} (cf. Eq.~(\ref{eq:hannamptangest})) and ``PN
evolved'' from \cite{HusHanGon07}. The right panel gives the deviation in
percent with respect to the 3PN-circular curve.}
\label{fig:ptangcompareadmtoeob}
\end{figure*}

\begin{figure*}[!ht] % 2-column figure 
\centering
\includegraphics[width=8.3cm,viewport=84 595 320 761]{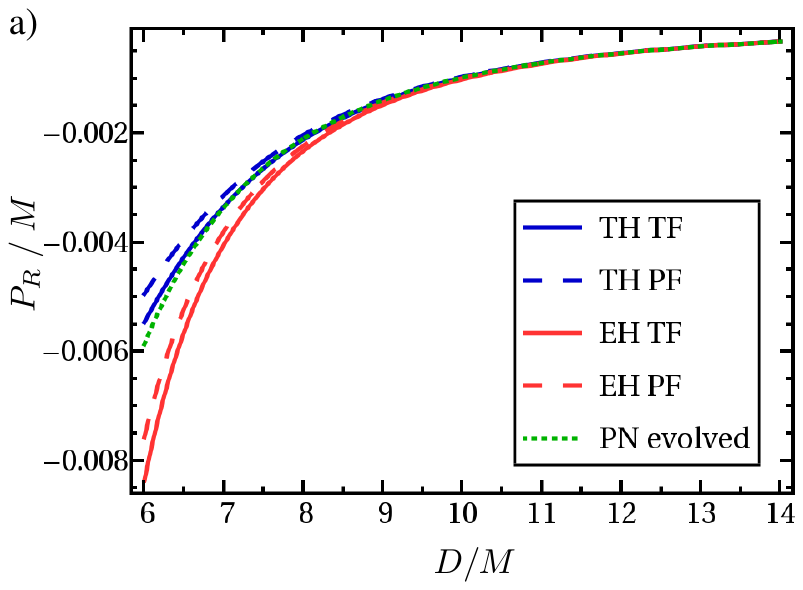}
\hskip0.8cm
\includegraphics[width=8.3cm,viewport=81 595 320
761]{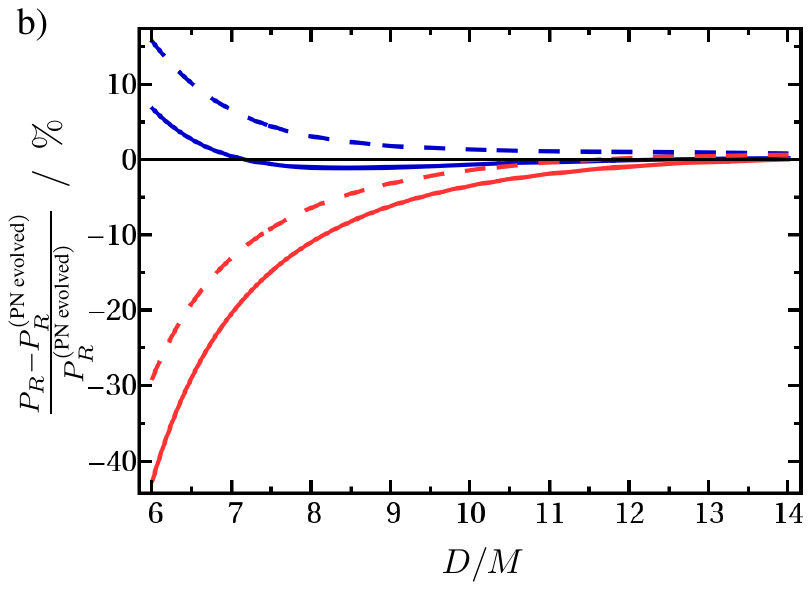}
\caption{(Color online) Radial momentum obtained with the initial data algorithm
for an 
equal-mass binary ($M=1$, no spins). Compared are combinations of the two 
Hamiltonians (Taylor or EOB) with the two possible flux functions (Taylor or 
Pad\'{e}). As expected, the THTF curve almost coincides with the reference 
function ``PN evolved'' from \cite{HusHanGon07}.}
\label{fig:pradcompareadmtoeob}
\end{figure*}

\begin{figure*}[!ht] % 2-column figure 
\centering
\vskip0.3cm
\includegraphics[width=8.3cm,viewport=84 595 320 772]{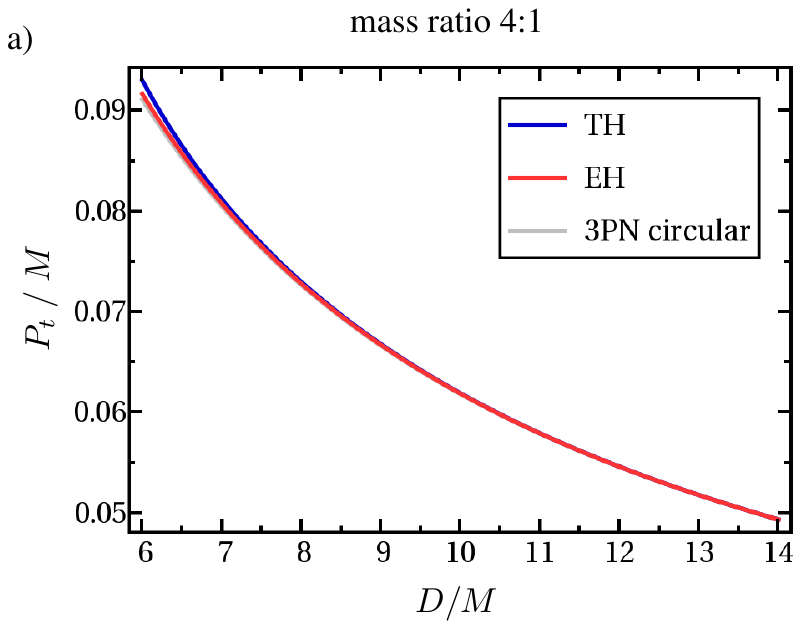}
\hskip0.8cm
\includegraphics[width=8.3cm,viewport=84 595 320 772]{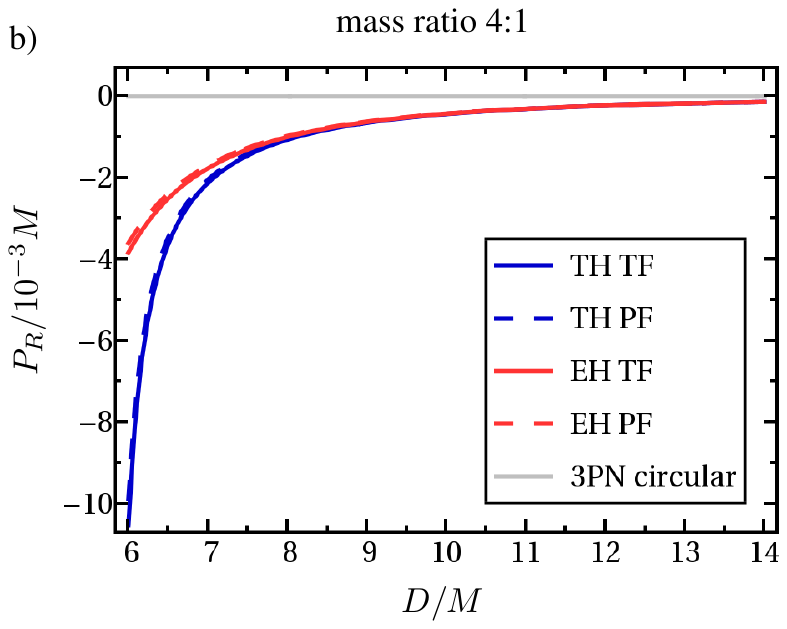}
\\[1.0cm]
\includegraphics[width=8.3cm,viewport=84 595 320 772]{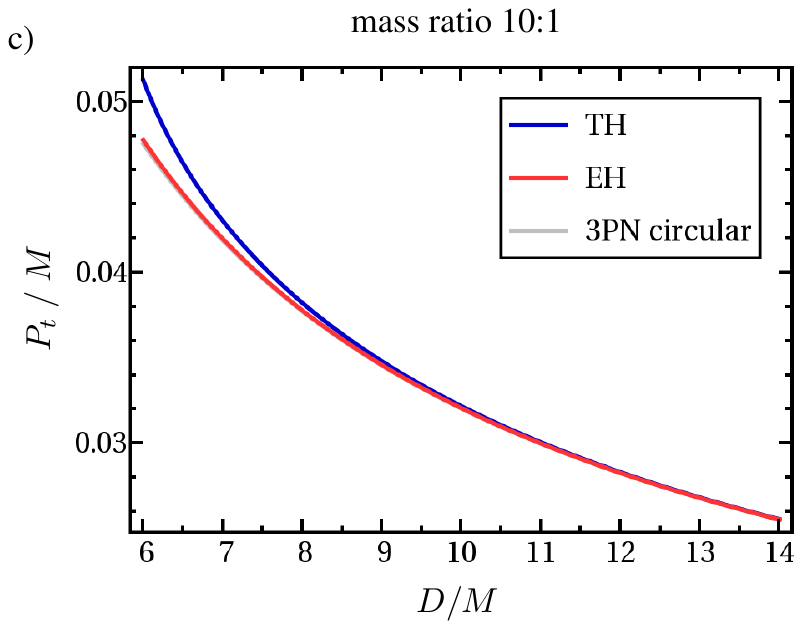}
\hskip0.8cm
\includegraphics[width=8.3cm,viewport=84 595 320 772]{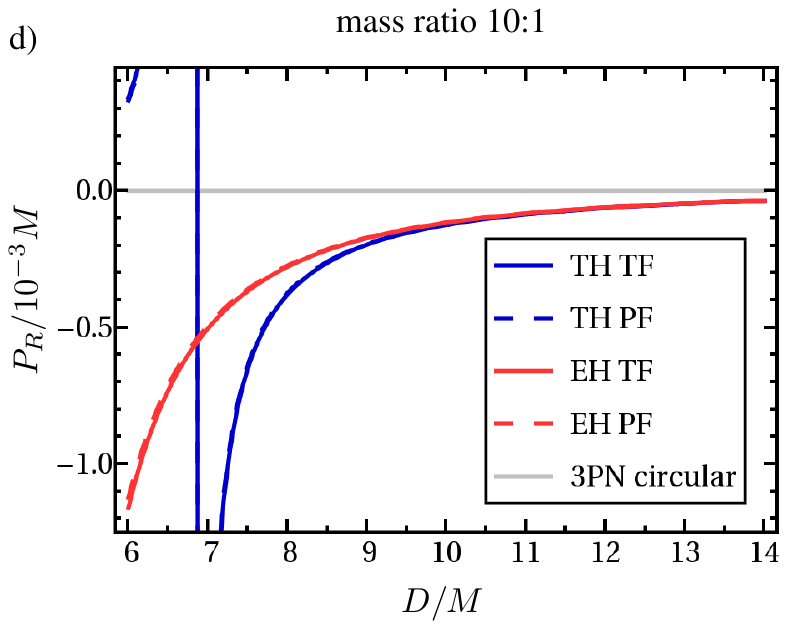}
\caption{(Color online) Behavior of Taylor-expanded and EOB initial data for
higher 
mass ratios ($M=1$, no spins). For $P_t$, EH and ``3PN circular'' almost
coincide, while $P_R=0$ for ``3PN circular''.}
\label{fig:pradcompareadmtoeobmassratios}
\end{figure*}

Translating EOB results to ADM coordinates involves a certain degree
of inaccuracy because the transformation is only known approximately
(to 3PN order). Table~\ref{tab:relativeerrorinmomentum} gives some
conservative, rough error estimates based on PN convergence as shown in
Fig.~\ref{fig:eobvsadmcoordinates}. The estimates already
include the error contribution that would arise from matching with a
given ADM separation when applying the shooting procedure. Apparently,
the transformation can be regarded more accurate for larger
separations and higher mass ratios.
We conclude this section by presenting some examples for the orbital
parameters $P_t$ and $P_R$ obtained by the algorithm.
Figs.~\ref{fig:ptangcompareadmtoeob} and
\ref{fig:pradcompareadmtoeob} show initial parameters for an equal-mass
binary without spins. Significant deviations between the various
combinations, that is combining one of the Hamiltonians, either
Taylor-expanded (TH) or EOB Hamiltonian (EH), with one version of the
flux function, either Taylor-expanded (TF) or Pade-resummed (PF), can
be observed only at small separations. 
Some dependence on the choice of the flux function is observed in
Fig.~\ref{fig:pradcompareadmtoeob}, although the result depends more
strongly on the choice of Hamiltonian.
This effect becomes more pronounced for higher mass ratios, see
Fig.~\ref{fig:pradcompareadmtoeobmassratios}. 
The plots clearly indicate the expected degradation of the TH parameters
for small separations, which is caused by a zero transition of the
second derivative $\partial^2 H/\partial R^2$ in
Eq.~(\ref{eq:debydrsph}), while the first derivative and thus
$\partial^2H/(\partial R \partial L)$ does not vanish. The resulting
pole moves outwards to larger separations and spoils the data as the
mass ratio increases. We would conclude that the accompanying drift in the
$P_t$ curves is also a result of the Taylor Hamiltonian's behavior. Clearly, it 
would not make
sense to use TH data in the vicinity of the pole, Fig.~\ref{fig:pradcompareadmtoeobmassratios}\,d.

The EOB data on the other hand appear sound and remain close to the
$P_t$ values of~\cite{BruGonHan06} (our Eq.~(\ref{eq:hannamptangest})),
but might be affected by the transformation error. In fact, in the
equal-mass case there is concern that the transformation error exceeds
the difference to the TH data. However, since the error goes down at
higher mass ratios we expect the EOB data to be superior in these
cases.

We examine the data's behavior in the presence of spins at the end of
Section \ref{sec:numresults}.

%%%%%%%%%%%%%%%%%%%%%%%%%%%%%%%%%%%%%%%%%%%%%%%%%%%%%%%%%%%%%%%%%%%%%%%%%

\section{Numerical results}
\label{sec:numresults}

For a specific set of parameters, comprising masses, initial
separation and spins, the initial data algorithm can give us
tangential and radial momentum of a relative particle. The actual
initial configuration, consisting of the individual locations and
momenta of the black holes, can be obtained by placing the bodies
appropriately in a coordinate frame by setting
$\bs{P}_1=-\bs{P}_2=\bs{P}$ and e.g.\ letting the center of mass
coincide with the origin. We are also free to apply rotations.

As discussed the input parameters $m_a,
\bs{X}_a,\bs{P}_a, \bs{S}_a$ are formulated in (almost) ADMTT gauge
whereas the BAM code uses a different spatial metric
$\gamma_{ij}=\psi_0^4\,\tilde{\gamma}_{ij}$ with a conformally flat
background metric $\tilde{\gamma}_{ij}=\delta_{ij}$ on a maximal slice
($K=0$) (so-called black hole puncture data~\cite{BraBru97}).
In previous work it was found numerically that the two gauges are rather
close to each other (e.g.\ \cite{BruGonHan06,HusHanGon07}). 
However, it is by no means trivial nor entirely understood why
PN-non-eccentric initial parameters should produce non-eccentric
orbits in full GR.
First, the process of translating the initial parameters to
the actual initial data involves the solution of the Hamiltonian
constraint, which, for example, alters the mass parameters,
cf.\ Table~\ref{tab:initialdataconstraints}.
Second, the modelled configuration lacks the gravitational-wave
content which would be there if the binary had reached this
state in a physically realistic way. 
The constraint solution process and the ad-hoc specification of data
at a given time result in artificial waves in the initial data, which 
e.g.\ affects the black-hole positions in the beginning of the simulation.
Third, the gauge (lapse and shift) evolves in the starting
phase, settling down to the final values after about $50M$ of
coordinate time.
These effects may not only outweigh the hard-won subtle differences
between THTF, THPF, EHTF and EHPF initial data, but also make it
difficult to reliably assess the quality of these data.
\begin{table}[ht]
\begin{ruledtabular}
\begin{tabular}{ccP{1.4}}
\multicolumn{1}{c}{\rule{0pt}{11pt}Mass ratio} &
\multicolumn{1}{c}{$h_{\text{min},i} / M_i$} &
\multicolumn{1}{c}{$r_\text{out} / M \phantom{0}$}\\
\hline
\rule{0pt}{11pt}
1:1 & $3/64$ & 774  \tn
%
%\hline
%
\rule{0pt}{11pt}
1:2 & $3/56$ & 2067 \tn
%
%\hline
\rule{0pt}{11pt}
1:4 & $1/96$ &  1236\tn
\end{tabular}
\end{ruledtabular}
\caption{
Grid setups of the different types of numerical simulations. The index
$i$ enumerates the black holes ($i=1,2$) and $M=M_1+M_2$ is the total
mass of the system.  $h_{\text{min},i}$
is the resolution of the finest grid covering the $i^\text{th}$
black hole. $r_\text{out}$ is the position of the outer boundary. All
results throughout the paper corresponding to the same mass ratio (but
e.g.\ different initial separations or spins) were obtained with the
setup given here.  }
\label{tab:numsetup}
\end{table}

\begin{table*}[!ht]
\centering
\begin{tabular}{x{2cm}x{1.8cm}x{2.2cm}P{1.6}P{2.6}x{0.5cm}x{2.0cm}x{0.5cm}P{1.4}}
\hline
\hline
\multicolumn{1}{c}{\rule{0pt}{11pt}Mass ratio} & \multicolumn{1}{c}{$D~/~M$} & \multicolumn{1}{c}{Data type} & \multicolumn{1}{c}{$~~P_t~/~M~~~~~~$}& \multicolumn{1}{c}{$P_R~/~10^{-3}M$}& &\multicolumn{1}{c}{Interval $/M$} & &\multicolumn{1}{c}{Eccentricity}\\
\hline
\rule{0pt}{11pt}1:1 & 11 & THTF &   0.090110 &  -0.71442 & &300$\,\ldots\,$600 && 0.002\phantom{0} \\
    &    & EHPF &   0.090266 &  -0.71489 & &300$\,\ldots\,$600 && 0.003\phantom{0} \\
\hline
1:1 &  7 & THTF &   0.123863 &  -3.34124 &  &\phantom{0}60$\,\ldots\,$120 && 0.005\phantom{0} \\
    &    & THPF & \multicolumn{1}{c}{$~~''$} & -3.13589 & &\phantom{0}60$\,\ldots\,$120 && 0.005\phantom{0}\\
    &    & EHTF &   0.125288 &  -4.04062  &  &\phantom{0}80$\,\ldots\,$150 && 0.005\phantom{0}\\
    &    & EHPF & \multicolumn{1}{c}{$~~''$} & -3.79255 & &\phantom{0}80$\,\ldots\,$150 && 0.005\phantom{0}\\
\hline
2:1 & 10 & THTF &   0.085615 &  -0.79486 & &250$\,\ldots\,$400 && 0.002\phantom{0} \\
    &    & EHTF &   0.085753 &  -0.80701 & &250$\,\ldots\,$400 && 0.003\phantom{0} \\
\hline
4:1 & 10 & THTF &   0.061921 &  -0.43393 & &300$\,\ldots\,$700 && 0.004\phantom{0}\\
    &    & EHPF &   0.061883 &  -0.42106 & &300$\,\ldots\,$700 && 0.0025\rule[-4pt]{0pt}{11pt}\\
\hline
\hline
\end{tabular}
\caption{
Simulations of black hole binaries without spin for equal and unequal
masses. A local eccentricity measurement was performed in the time
interval given. 
The uncertainty in the eccentricity is estimated to be $\pm0.001$.  }
\label{tab:initialdata}
\end{table*}

For the purpose of the latter we investigated several approaches for
measuring the eccentricity of the inspiralling binary's orbit. Following the
ideas of \cite{BuoCooPre06,BoyAlt07,HusHanGon07,PfeAlt07}, we employ
two approaches based on the evolution of the separation $D(t)$ and the orbital
frequency $\omega(t)$. The first consists of comparing the actual with an ideal
(quasi-circular) inspiral curve $D_c(t)$ or $\omega_c(t)$, which is obtained by
averaging over the eccentricity oscillations using an appropriate fit function
(also see Fig.~\ref{fig:chi85orbits}). 
The resulting eccentricities are referred to as $e_D$ and $e_\omega$. 
Specifically, we employ a
model function according to $D_c(t)=\sum_{i=1}^{4}a_i\,(t_M-t)^{i/2}$ and
eccentricity is computed as $e_D(t)=[D(t)-D_c(t)]/D_c(t)$ \cite{HusHanGon07}.
The time $t_M$ entering the formula for the ideal separation $D_c(t)$ is the
merger time which in our case is estimated by the fitting procedure. 
The resulting value agrees very well with results from different
methods for measuring the merger time, for example the time when the
amplitude of $\Psi_4$ reaches its maximum.
As a model function for the frequency we use a fourth order polynomial,
$\omega_c(t)=\sum_{i=0}^{4}\,b_it^i$, following \cite{BakMetMcW06a}. In this
case, the eccentricity is calculated via
$e_\omega(t)=[\omega(t)-\omega_c(t)]/2\omega_c(t)$.
The fit to the model function only works well if we exclude initial gauge
adjustments and it fails at merger time ($D_c(t=t_M)=0$ in the $D$-method and
the fourth order polynomial ceases to capture the behavior of the frequency in
the $\omega$-method). We therefore have to choose an appropriate time interval
for the averaging procedure. 
Figures~\ref{fig:um4eccentricity}--\ref{fig:spineccentricityenlarge} show
graphs for eccentricities varying with time resulting from the $D$-method 
according to the formula given above. 
The global extremum of each curve, usually located at early times, is then
used to determine a time-independent eccentricity value of the data
which will serve to compare different runs with each other.

A second method tries to capture the oscillations themselves by
presuming some functional dependence, i.e.\ a sinusoidal fluctuation
on a certain background function like
$\dot{D}(t)=A_1 + A_2 t + B\sin(\Omega_0 t + \phi_0)$, where $A_1, A_2, B,
\Omega_0$ and $\phi_0$ are fit parameters. 
This works best for the time derivatives $\dot{D}(t)$ and $\dot{\omega}(t)$
instead of the original functions.
The resulting eccentricity can be estimated to be
$e_{\dot{D}}(t)\approx-B\cos(\Omega_0 t + \phi_0)/\Omega_0 D(t)$.
The procedure is equally applicable to the derivative of the frequency curve,
$\dot{\omega}(t)$, and leads to an eccentricity denoted by
$e_{\dot{\omega}}(t)$.
\\
Since the assumed ``capture'' functions are only able to trace the curves
$\dot{D}(t)$ and $\dot{\omega}(t)$ locally (in a very short time interval),
eccentricities computed with this approach turned out to be less 
useful for our purpose. 

\newpage
All in all, the different methods ($e_D$, $e_\omega$,
$e_{\dot{D}}$ and $e_{\dot{\omega}}$) do not always give consistent
results for very low eccentricities, for which the quality of the
fit function becomes decisive. Eccentricity values given in the
present paper are based on $e_D$, with error estimates based on the
variations in the extrema of $e_D$.

%%%%%%%%%%%%%%%%%%%%%%%%%%%%%%%%%%%%%%%%%%%%%%%%%%%%%%%%%%%%%%%%%%%%%%%%%%%%%%%%
Numerical evolutions were performed with the BAM
code~\cite{BruGonHan06,BruTicJan03} which uses the BSSN formulation of
the Einstein equations~\cite{ShiNak95,BauSha98} and the method of
moving punctures~\cite{CamLouMar05,BakCenCho05}. Spatial derivatives
are 6$^\text{th}$ order accurate~\cite{HusGonHan07} and a
4$^\text{th}$ order Runge-Kutta scheme is employed for time
stepping. We use ``1+log'' slicing for determining the lapse
function~\cite{BonMasSei94} and the ``000''-version of the
$\tilde{\Gamma}$-driver shift condition~\cite{AlcBruDie02,GunGar06}.
In order to resolve the black hole regions as well as the wave zone
sufficiently well, mesh refinement with nested boxes is used. Details of
the grid setups used for numerical simulations are given in
Table~\ref{tab:numsetup}.  Note the choice of resolutions for the
black holes: the number of refinement boxes is chosen for each black
hole individually in order to obtain the same effective resolution
$h_\text{min}$ for both of them.
%%%%%%%%%%%%%%%%%%%%%%%%%%%%%%%%%%%%%%%%%%%%%%%%%%%%%%%%%%%%%%%%%%%%%%%%%%%
\subsection{Numerical results for different mass ratios and vanishing spin}
\begin{figure}[!ht] 
\centering
\includegraphics[width=8.3cm,viewport=84 595 320
767]{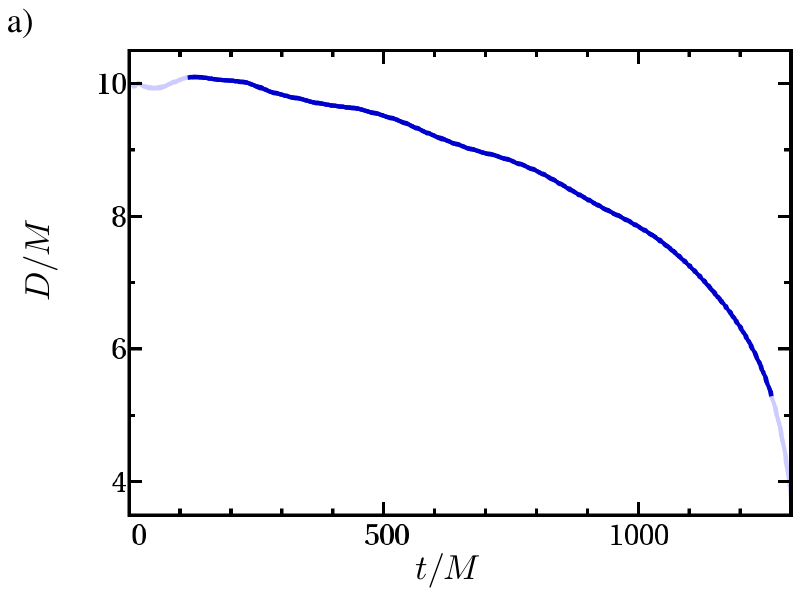}
\hskip0.8cm
\includegraphics[width=8.3cm,viewport=84 595 320
767]{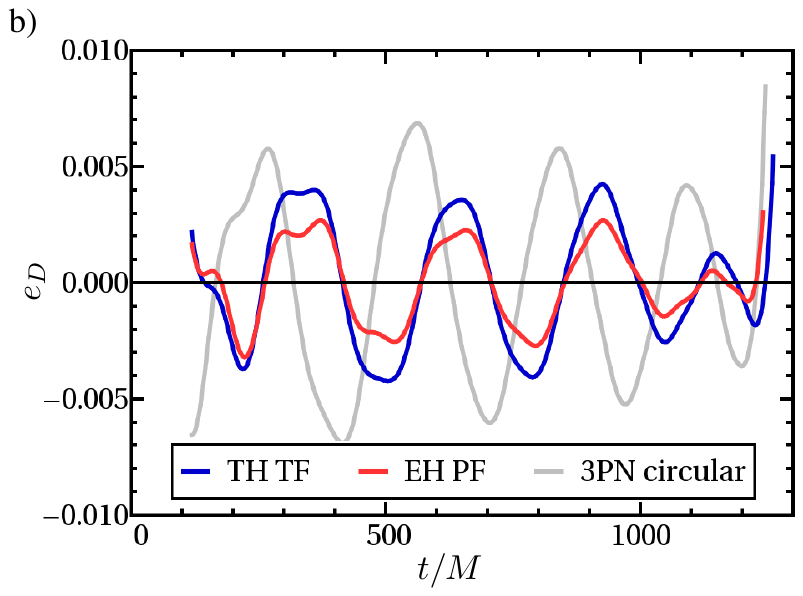}
\vskip-0.5cm
\caption{
(Color online) 
Dynamics and eccentricity for mass ratio 4:1 ($D=10M$). Panel (a)
shows the oscillating decrease of the binary separation for the THTF
data. The darker part of the line indicates the time interval used for
the fit required for $e_D$. In panel (b) the eccentricity
is compared between Taylor-expanded and EOB-Hamiltonian data.
}
\label{fig:um4eccentricity}
\end{figure}
\begin{figure}[!ht]
\centering
\includegraphics[width=8.3cm,viewport=164 363 400
561]{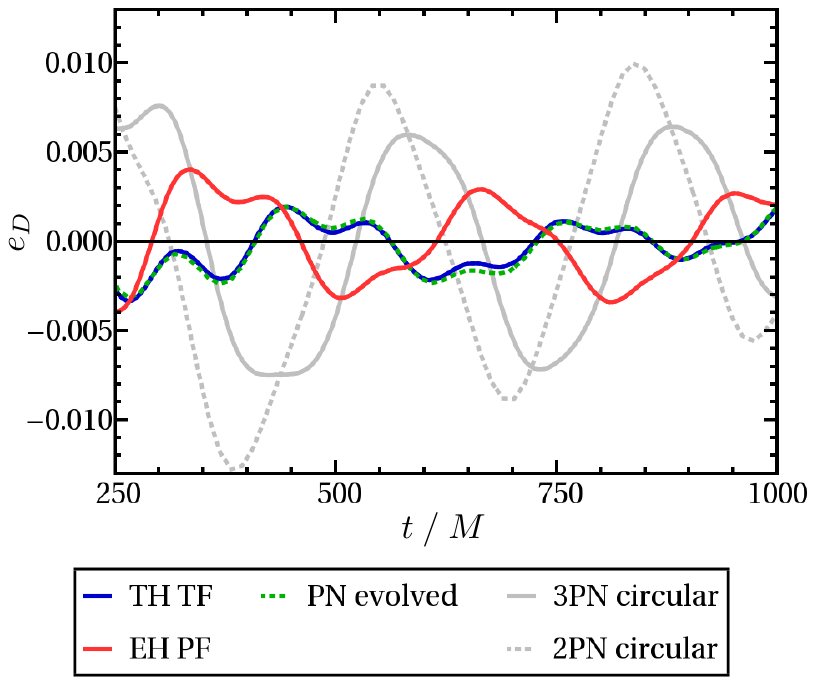}
\caption{
(Color online) 
Eccentricity for mass ratio 1:1 ($D=11M$). Compared are runs for PN-circular 
data, PN-evolved data,
and the new THTF and EHPF data. The THTF run is very similar to the PN-evolved 
run as expected, while for
2PN circular data the eccentricity is up to a factor of five larger.
}
\label{fig:emeccentricity}
\end{figure}
Table \ref{tab:initialdata} summarizes the outcome of eccentricity
measurement for various simulations with equal and unequal masses
without spin. The initial separation was partially chosen to allow
direct comparison with~\cite{HusHanGon07,DamNagHan08}.
Fig.~\ref{fig:um4eccentricity} shows the time dependence of $e_D$ for 
mass ratio 4:1, giving a visual impression of the 
oscillations of the inspiralling orbit, while
Fig.~\ref{fig:emeccentricity} shows results for 1:1.

We draw the following conclusions from the numerical results.
Our techniques for measuring eccentricity give compatible results
in most cases, but in some cases the accuracy of $\pm0.001$ is not
quite sufficient to reliably resolve the smallish differences. Unless
the eccentricities deviate by amounts larger than the uncertainties we
cannot name one of the data the better choice.
In particular, we examined the consequences of the choice of flux for
the smallest initial separation $D=7M$ case, where the differences
should be largest. At the given level of accuracy 
we cannot discern a significant difference between the
TF and PF methods.

For initial separations of $D=10M$ and $D=11M$, the eccentricity
ranges from $e=0.002$ to $0.004$ for comparatively small variations of
about 1\% in $P_t$ and 2\% in $P_R$. This reflects the high
sensitivity of the eccentricity to small changes in the momenta. The
minimal eccentricity obtainable by the method is about $e=0.0025$ for
mass ratios from 1:1 to 4:1. 

For smaller initial binary separation the
PN initial data become less appropriate since the underlying equations
approach their limit of validity. Moreover, the deviation in the
gauges, that is ADMTT gauge on the PN side versus maximal slicing in
the numerical simulation, grows. The numerical example for $D=7M$
shows that the eccentricity grows to $e=0.005$ for equal masses, 
which is twice as large as for $D=11M$. Considering how close $D=7M$
is to the merger, it is surprising how well the PN initial parameters
work in this context.

Going to higher mass ratios the putative advantage of EH data
first seems to show at 4:1, although the difference in initial
momentum to TH data is quite small. Given the available data, the
smaller eccentricity might be accidental, but we believe there
is a systematic trend based on corresponding results for spinning
binaries. (Numerical simulations at 10:1 have
recently become available~\cite{GonSpeBru08}, so the behavior for
increasing mass ratios could now be investigated further.)

Utilizing the same Hamiltonian, our TH data are expected to give
results similar to the PN-evolved data, and this is indeed the
case. For example, \cite{HusHanGon07} obtained $e_D=0.002\pm0.001$ for
the 1:1 case starting at $D=11M$.
Our TH results are also consistent with the outcome of PN-evolved
data at higher mass ratios, for example \cite{DamNagHan08} found
$e=0.003$ for mass ratio 2:1 and $e=0.005$ for the 4:1 case at
$D=10M$. 

In~\cite{HusHanGon07}, which introduced the PN evolution method to
obtain initial parameters with reduced eccentricity, the PN evolutions
were started with 3PN circular parameters of~\cite{BruGonHan06}, and
the result was compared to numerical simulations starting from 2PN
(one order lower) circular parameters based
on~\cite{Kid95,BruGonHan07}. The PN-evolved data showed an improvement
by a factor of up to five in the eccentricity compared to the 2PN
circular parameters for mass ratio 1:1.
In Fig.~\ref{fig:emeccentricity} we compare the result for 1:1 for 2PN and 3PN
circular parameters, the PN-evolved parameters, and the new THTF and
EHPF parameters.
The THTF run is very similar to the PN-evolved run as already pointed out. As
stated in Table~\ref{tab:initialdata}, the EHPF parameters result 
in about 50\% larger eccentricity (for equal masses).
For numerical simulations starting from the 3PN instead of 2PN
circular parameters the improvement is less, about a factor of two for
mass ratio 4:1 and 1:1 as shown in Figs.~\ref{fig:um4eccentricity}
and~\ref{fig:emeccentricity}. 
Incidentally, note that the 2PN circular parameters are based on
harmonic coordinates while the 3PN circular parameters refer to ADMTT.

Our final comment on the vanishing spin case is that, as shown in
Figs.~\ref{fig:um4eccentricity} and~\ref{fig:emeccentricity}, there
still is a distinctive orbital eccentricity for mass ratio 4:1, while
for 1:1 the eccentricity is reduced to a level where the oscillations
are less clearly associated with the orbital motion. At least for 4:1
there seems to be room for improvement.

%%%%%%%%%%%%%%%%%%%%%%%%%%%%%%%%%%%%%%%%%%%%%%%%%%%%%%%%%%%%%%%%%%%%%%%%%%

\subsection{Numerical results for equal masses and non-vanishing spin}

Let us now turn to spinning black hole binaries for equal masses.
We would like to emphasize that as soon as spins are present eccentricity is
inevitable. This is due to spin-spin interaction and can be seen from
the PN equations of motion (\ref{eq:spineom}), which predict a
precession of the spins with a frequency determined by
(\ref{eq:hss}). This precession influences the orbital
motion via Eq.~(\ref{eq:hso}). Moreover, we find an even more direct
impact on the orbital motion as a consequence of the $R$-dependence in
(\ref{eq:hss}) which directly affects the evolution of
$\bs{P}$ in (\ref{eq:nonconservp}). The latter mechanism is also
relevant when both spins are parallel to the orbital angular
momentum. 
Despite the constancy of $\bs{L}=\bs{X}\times\bs{P}$ in
these cases (ignoring radiation for the time being), the orbital
variables $\bs{X}$ and $\bs{P}$ would still be allowed to exhibit
the corresponding oscillations. 

Table~\ref{tab:spinecc}\,a illustrates this issue for an evolution
with the conservative PN equations. (This is different from the
PN-evolved parameters we discuss elsewhere which include flux terms.)
\begin{table*}[!ht]
\centering
\begin{tabular}{c}
a)~~\\
\rule{0pt}{14pt}\\
\\
\\
\\
\\
\\
\end{tabular}
\begin{tabular}{x{1.7cm}x{2.0cm}x{2.0cm}}
\hline
\hline
\multicolumn{1}{c}{$\chi$} & \multicolumn{2}{c}{$e_D$ (PN evolution)}\tn 
& $D=10M$ & $D=8M$ \tn
\hline
\rule{0pt}{12pt}-0.25 & 0.0020& 0.0040\tn
\phantom{-}0.00 & 1$\,\cdot\,$10$^\text{-7}$ & 3$\,\cdot\,$10$^\text{-7}$\tn
\phantom{-}0.25 & 0.0015 & 0.0024 \tn
\phantom{-}0.50 & 0.0050 & 0.0083\tn
\phantom{-}0.85 & 0.014\phantom{0} & 0.022\phantom{0}\tn
\hline
\hline
\end{tabular}\hskip0.8cm
\begin{tabular}{c}
b)~~\\
\rule{0pt}{14pt}\\
\\
\\
\\
\\
\\
\end{tabular}
\begin{tabular}{x{1.7cm}x{3.0cm}x{3.0cm}}
\hline
\hline
\multicolumn{1}{c}{$\chi$} & $e_D$ (TH)& $e_D$ (EH)\tn 
& & \\
\hline
\rule{0pt}{12pt}-0.25     & 0.015$\,\pm\,$0.002   & 0.005\phantom{0}$\,\pm\,$0.001\phantom{0}\tn
\phantom{-}0.00 & 0.002$\,\pm\,$0.001   & 0.003\phantom{0}$\,\pm\,$0.001\phantom{0}\tn
\phantom{-}0.25           & 0.015$\,\pm\,$0.003   & 0.0015$\,\pm\,$0.0005\tn
\phantom{-}0.50 & 0.034$\,\pm\,$0.005   & 0.005\phantom{0}$\,\pm\,$0.001\phantom{0}\tn
\phantom{-}0.85           & 0.07\phantom{0}$\,\pm\,$0.01\phantom{0}     & 0.014\phantom{0}$\,\pm\,$0.003\phantom{0}\tn
\hline
\hline
\end{tabular} 
\caption{
a) Eccentricities from conservative PN evolution for aligned equal
spins at $D=10M$ and $D=8M$. 
b) Results from corresponding simulations with BAM at comparable
binary separations, cf. Fig.~\ref{fig:spineccentricity}.}
\label{tab:spinecc}\vspace{1cm}
\end{table*}%
\begin{table*}[!ht]
\centering
\begin{tabular}{x{1.5cm}x{1.0cm}x{1.3cm}x{1.8cm}P{1.6}P{2.6}x{0.1cm}x{1.0cm}x{1.0cm}x{1.3cm}x{1.3cm}}
\hline
\hline
\multicolumn{1}{c}{\rule{0pt}{14pt}Mass ratio} &\multicolumn{1}{c}{$\chi$}& \multicolumn{1}{c}{$D~/~M$} & \multicolumn{1}{c}{Data type} & \multicolumn{1}{c}{$~~P_t~/~M~~$}& \multicolumn{1}{c}{$P_R~/~10^{-3}M$}& &\multicolumn{1}{c}{$m_1$} & \multicolumn{1}{c}{$m_2$}& \multicolumn{1}{c}{$\tilde{m}_1$}& \multicolumn{1}{c}{$\tilde{m}_2$}\tn
\hline
\rule{0pt}{12pt}1:1 & -0.25 & 12 & THTF & 0.086803 & -0.60477 & &0.5 & 0.5 & 0.4757 & 0.4757\tn
        &       &    & EHPF & 0.086459 & -0.57913 & &0.5 & 0.5 & 0.4757 & 0.4757\tn
\hline
1:1 & \phantom{-}0.00 & 11 & THTF &   0.090110 &  -0.71442 & & 0.5 & 0.5 & 0.4872 & 0.4872\tn
    &  &    & EHPF &   0.090266 &  -0.71489 &  &  0.5  & 0.5  &  0.4872 &  0.4872 \tn
\hline
1:1 & \phantom{-}0.25  & 12 & THPF & 0.083322 & -0.47391 & &0.5 & 0.5 & 0.4758 & 0.4758\tn
        &       &   & EHPF & 0.083858 & -0.49460 & &0.5 & 0.5 & 0.4758 & 0.4758\tn
\hline
1:1 & \phantom{-}0.50  & 11 & THPF & 0.085921 & -0.54782 & &0.5 & 0.5 & 0.4328 & 0.4328\tn
        &       &   & EHPF & 0.087194 & -0.60170 & &0.5 & 0.5 & 0.4328 & 0.4328\tn
\hline
1:1 & \phantom{-}0.85  & 10 & THPF & 0.087364 & -0.60110 & &0.5 & 0.5 & 0.2564 & 0.2564\tn
& &  & EHPF& 0.090080 & -0.71349 & &0.5 & 0.5 & 0.2563 & 0.2563\tn
\hline
4:1 & \phantom{-}0.00 &  10 & THTF &   0.061921 &  -0.43393 &  & 2.0   & 0.5  &  1.9788 & 0.4761 \tn
 \rule[-5pt]{0pt}{12pt}   &  &    & EHPF &   0.061883 &  -0.42106 &  & 2.0   & 0.5  &  1.9788 & 0.4761\tn
\hline
\hline
\end{tabular}
\caption{
Initial parameters used in the simulations with spin for equal
masses. The same quantities are shown for a 4:1 run for
comparison. The total mass is $M=m_1+m_2$.}
\label{tab:initialdataconstraints}
\end{table*}
The test case consists of an equal-mass binary with spin settings
according to $\bs{S}_i=Gm_i/c^2\cdot\chi\,\hat{\bs{L}}$. We applied
suitable initial data from the above algorithm which would give
circular orbits if only $H_{\text{SO}}$ was involved (in particular we
dropped the $[P_R]_0$ component). 
The observed radial fluctuations are of orbital frequency (when taking
periastron advance into account) and can be explained by means of
Eqns.~(\ref{eq:hss})-(\ref{eq:hsisi}). Since neither $\bs{L}$ nor the
spins precess in the current constellation, the spin-spin contribution
just represents a spherically symmetric deformation term
$H_\text{SS}=\text{const}\cdot R^{-3}$. Due to its simplicity in this
particular case the correction could in principle be taken into
account when computing the initial parameters. For general spin
configurations, however, this is not advisable, so we do not pursue
this idea here but return to it later.  Instead we can regard the
numbers in Table~\ref{tab:spinecc}\,a as an estimate of the minimal
eccentricities to be achieved in fully numerical simulations.
(For $\chi=0$, the eccentricity is analytically zero, and
the small deviations from zero in Table~\ref{tab:spinecc}\,a are
numerical error of the integration method.)

The aligned and anti-aligned scenarios were explored with several runs.
The main purpose was to find out which of the data type
(THTF, THPF, EHTF, EHPF) leads to minimal eccentricity. The observation
that the choice of the flux does not play an important role seems to
carry over to the spinning case, so we focussed on the effect of the
Hamiltonian. Starting evolutions at separations between
$D=10\ldots12M$ we varied the spin
magnitude. 
Table~\ref{tab:initialdataconstraints} gives the corresponding initial
parameters and also shows the effect of constraint solving. For
spinning holes the bare masses $\tilde{m}_1$ and $\tilde{m}_2$ are
reduced to keep the total energy constant.
In our notation $m_a$ replaces the capital $M_a$ often used in the
numerical literature in order to maintain the PN notation.

\begin{figure*}[!ht] % 2-column figure 
\centering
\includegraphics[width=8.3cm,viewport=84 595 320
768]{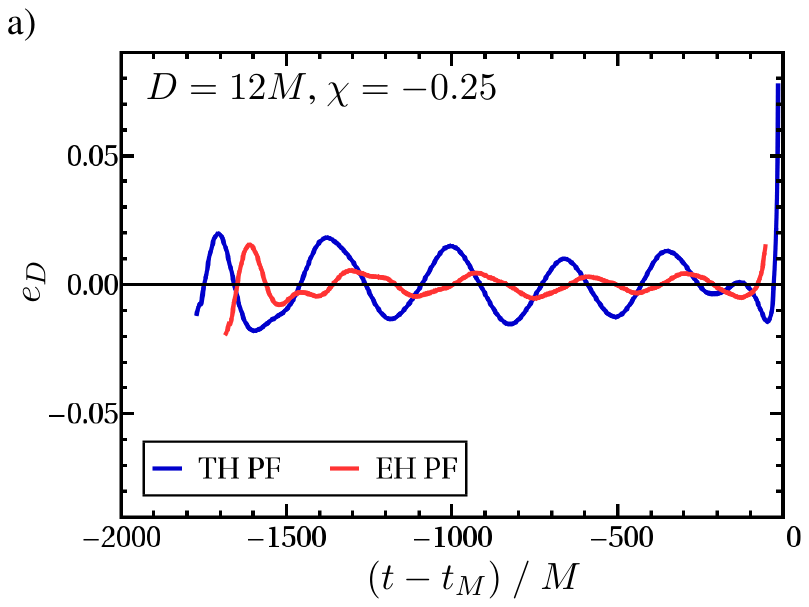}
\hskip0.4cm
\includegraphics[width=8.3cm,viewport=84 595 320
768]{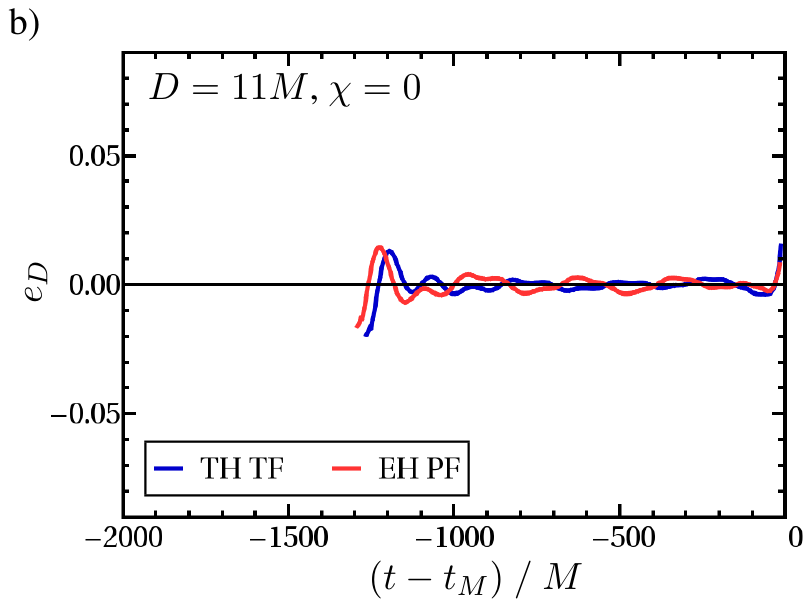}
\\[0.8cm]
\includegraphics[width=8.3cm,viewport=84 595 320
768]{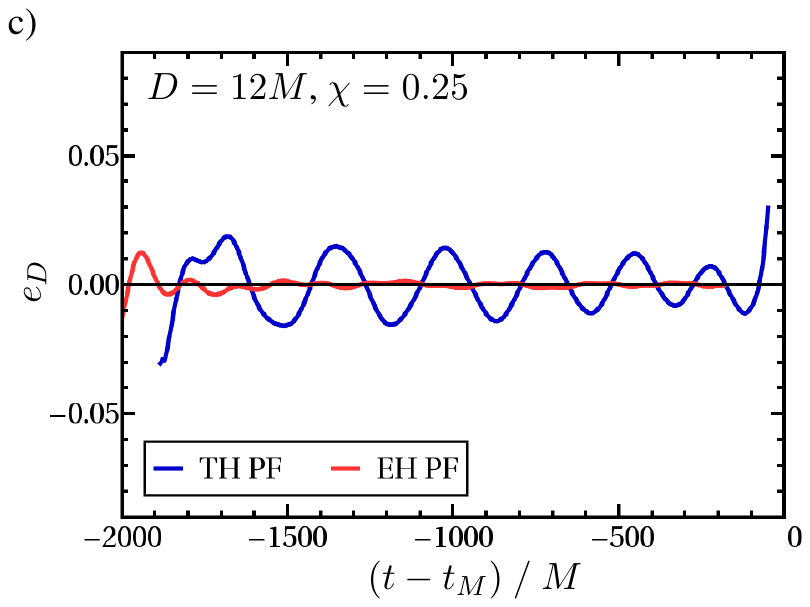}
\hskip0.4cm
\includegraphics[width=8.3cm,viewport=84 595 320
768]{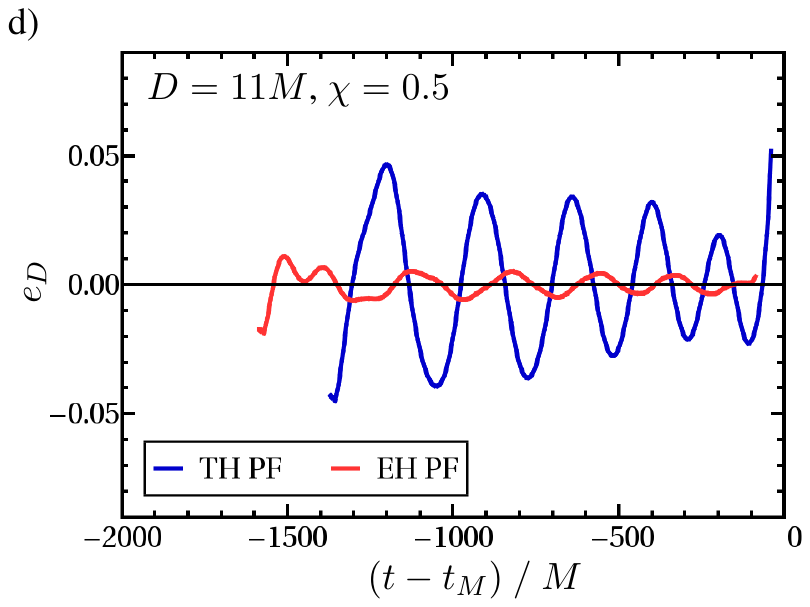}
\\[0.8cm]
\includegraphics[width=8.3cm,viewport=84 595 320
768]{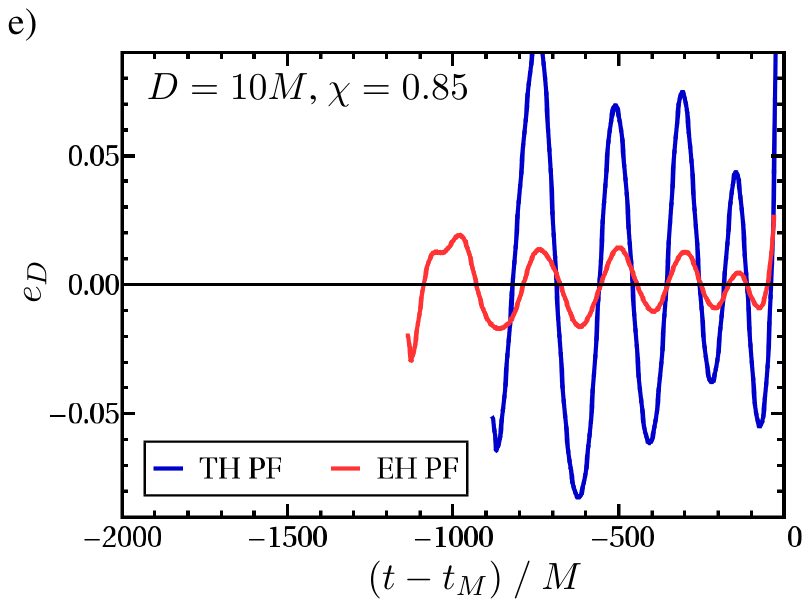}
\hskip0.4cm
\includegraphics[width=8.3cm,viewport=84 595 320
768]{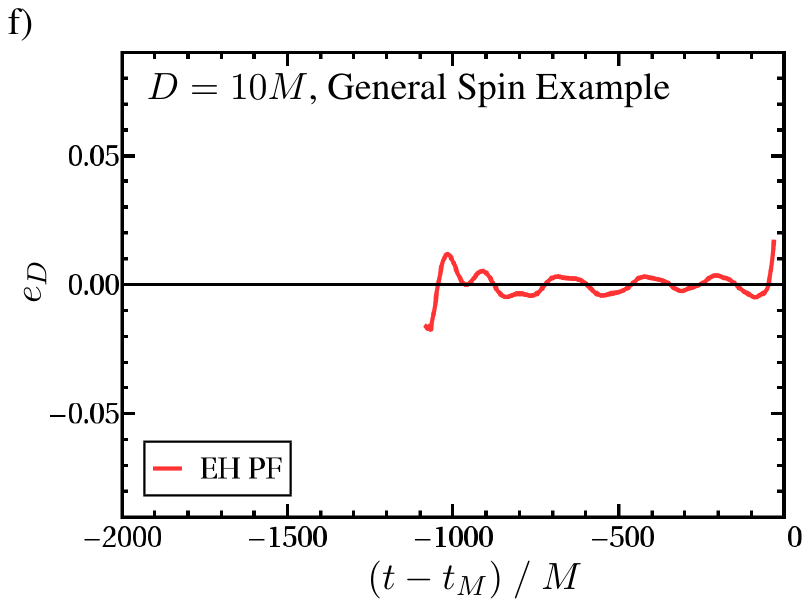}
\\[0.8cm]
\caption{
(Color online) 
Performance of PN initial data with spin in numerical
simulations. In the present equal-mass configurations ($m_1=m_2=0.5$),
both spin vectors point in the same direction parallel to
the orbital angular momentum and have the same length, i.e.\
$\bs{S}_i=Gm_i/c^2\cdot\chi\,\hat{\bs{L}}$ for panels (a) - (e). We
compare the eccentricity $e_D$ between TH and EH data for various
values of $\chi$. In panel (f) we set up a more general spin
configuration with
$\bs{S}_1=m_1^2\,\chi_1\,(0,-\cos\,45^\circ,\cos\,45^\circ)$ and
$\bs{S}_2=m_2^2\,\chi_2\,(-\cos\,60^\circ,0,\cos\,30^\circ)$ with
$\chi_1=\chi_2=0.5$.  In order to simplify comparisons of the plots, all
axes are scaled the same way, and $t_M$ denotes the moment of
merger.
Large amplitudes in the beginning of the graphs are due to gauge
adjustments in the code and should be ignored, while strong drifts
near the end of some curves arise when the fit function (needed to
compute $e_D$) becomes inappropriate.
}
\label{fig:spineccentricity}
\end{figure*}

\begin{figure*}[!ht] % 2-column figure 
\centering
\includegraphics[width=8.3cm,viewport=84 576 320
760]{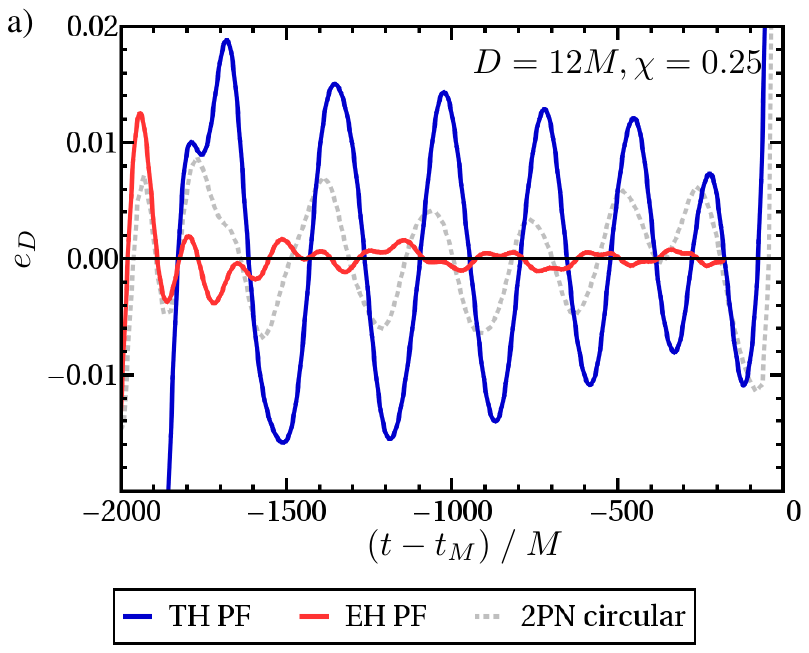}
\hskip1.0cm%
\includegraphics[width=8.3cm,viewport=84 576 320
760]{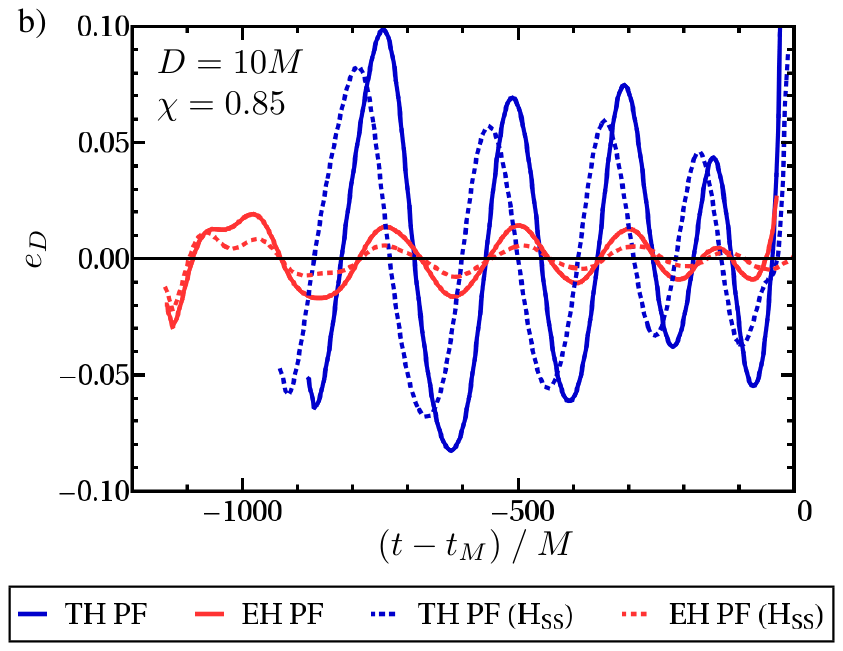}
\vskip0.5cm
\caption{
(Color online) 
Zoom-in plots for some of the cases of
Fig.~\ref{fig:spineccentricity} comparing to eccentricities obtained
with other methods.
a) For equal masses and spin $\chi=0.25$, 2PN-circular
data ($P_R=0$) amended with spin (see~\cite{HanHusBru07}) happen to
give larger eccentricity than the EH data with properly chosen $P_R$.
b) For equal masses and spin $\chi=0.85$, the benefits of including
spin-spin interactions in the initial data computation are shown. 
}
\label{fig:spineccentricityenlarge}\vspace*{0.0cm}
\end{figure*}

\begin{figure*}[!h] % 2-column figure 
\centering
\includegraphics[height=6.4cm,viewport=84 576 274
759]{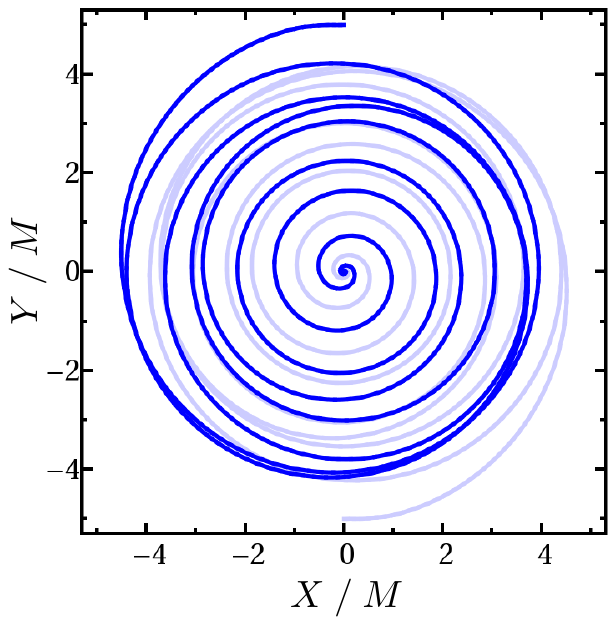}
\hskip0.4cm
\includegraphics[height=6.4cm,viewport=84 577 338
766]{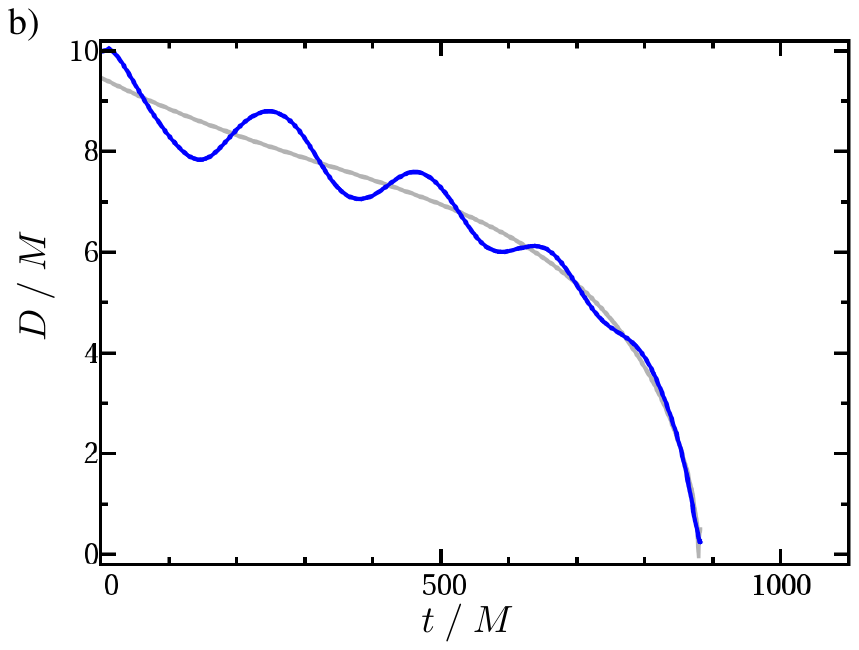}
\\[0.2cm]
\includegraphics[height=6.4cm,viewport=84 576 274
759]{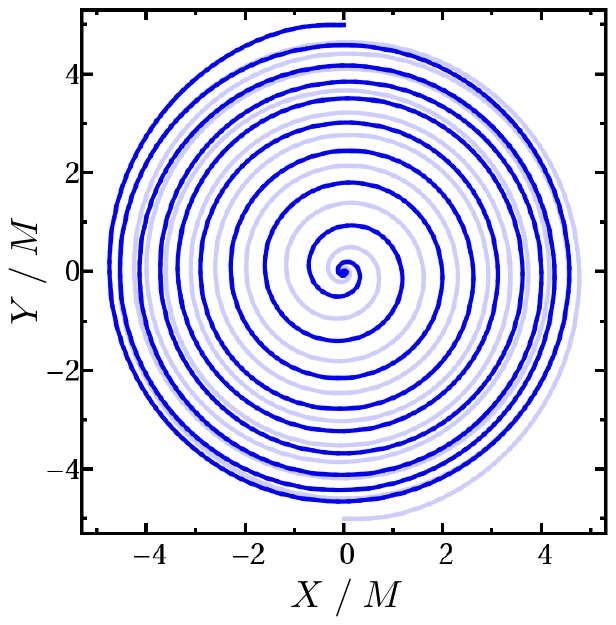}
\hskip0.4cm
\includegraphics[height=6.4cm,viewport=84 577 338
766]{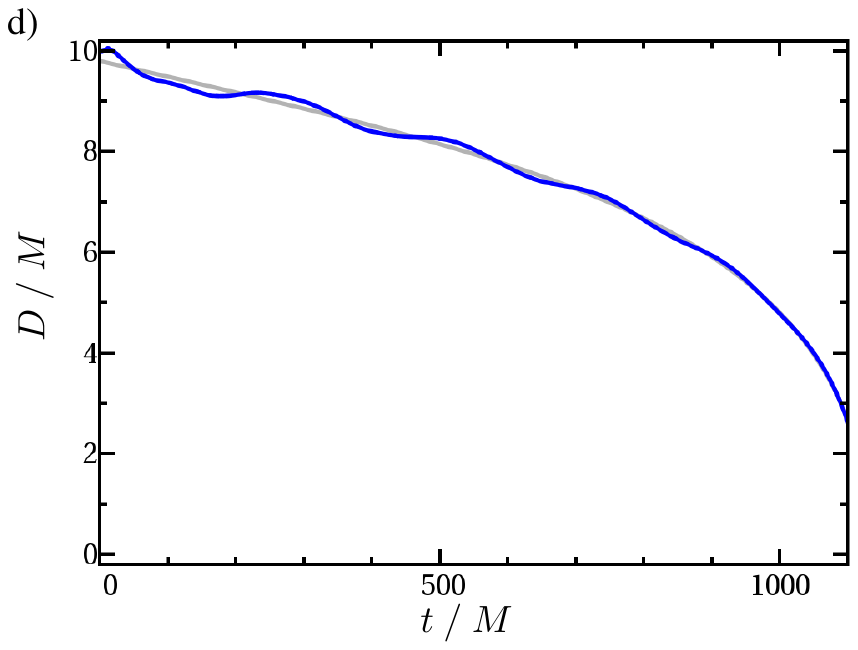}
\vskip-0.4cm
\caption{ 
(Color online) 
Puncture tracks (left panels) and puncture distance (right panels) for
the $\chi=0.85$ simulation from
Fig.~\ref{fig:spineccentricity}\,e. The upper plots pertain to the
Taylor-expanded data (THPF) and give a large eccentricity of
$e_D=0.07$, whereas the EOB data (EHPF) yield much smaller
fluctuations in the binary separation leading to $e_D=0.014$ (lower
plots). The right-hand-side graphs also indicate the procedure of
eccentricity measurement. The grey/lighter curve is the fit function
$D_c(t)$ representing an ideal, non-eccentric inspiral.
}
\label{fig:chi85orbits}
\end{figure*}
\newpage
Fig.~\ref{fig:spineccentricity} depicts the resulting eccentricity
graphs while Table~\ref{tab:spinecc}\,b provides the corresponding
numbers. 
Fig.~\ref{fig:spineccentricityenlarge} shows additional details for
the data in Fig.~\ref{fig:spineccentricity}, including results
obtained by other methods.

The main observation is that the eccentricity increases
significantly as $\chi$ is changed from zero towards positive and
negative values. For TH data, the minimum is near $\chi=0$, while for
EH data the minimum is closer to $\chi=0.25$ than $\chi=0$. 
The EOB data yield remarkably low eccentricities, and for $\chi\geq0.25$
the values are very close to the PN-limit estimate of
Table~\ref{tab:spinecc}\,a. 
The TH data perform poorly in comparison, with up to five times larger
eccentricity for $|\chi|\geq0.25$.
It would be interesting to extend this study to more negative values of $\chi$.
In order to illustrate these results Fig.~\ref{fig:chi85orbits} shows
the black-hole orbits of the simulation with $\chi=0.85$. Strong
deviations from a spiral can be observed for the Taylor-Hamiltonian
data while the EOB data produce a rather well-formed inspiral. Note
that the motion takes place in the $z=0$ plane.

\begin{figure*}[b] % 2-column figure 
\centering
\includegraphics[width=14.5cm,viewport=76 580 496
760]{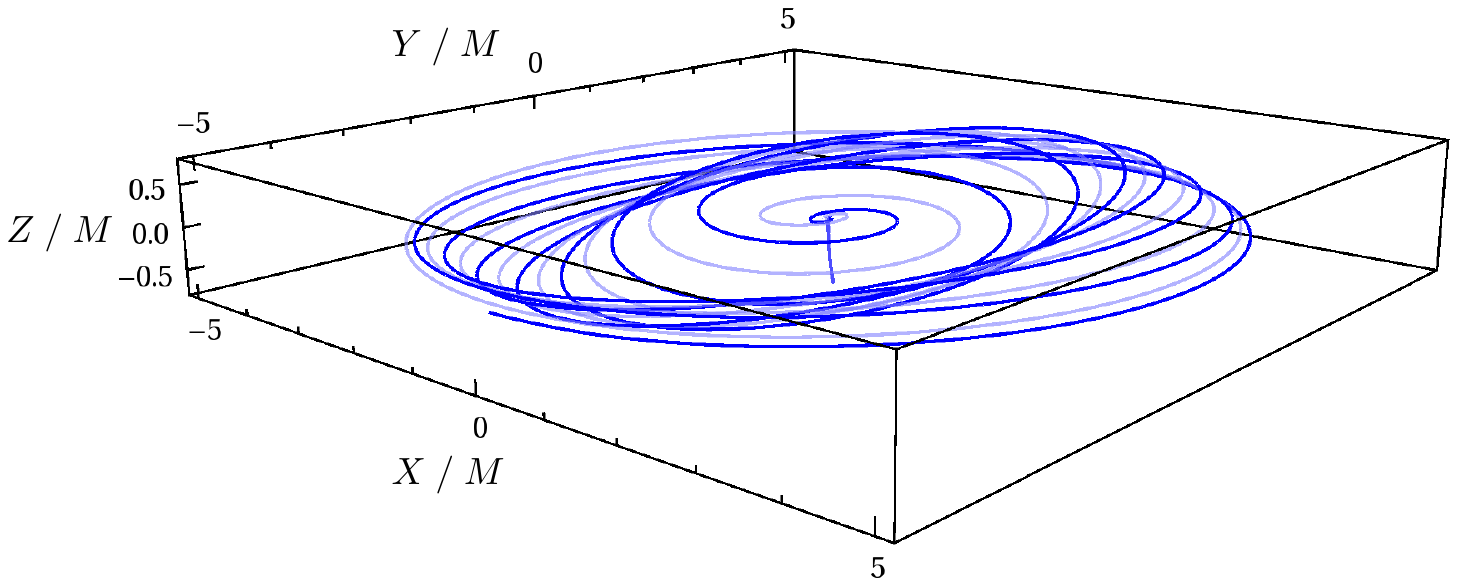}
\caption{(Color online) Orbital motion for the general spin example from 
Fig.~\ref{fig:spineccentricity}\,f. The orbital plane precesses and we observe a
spin kick of the final black hole.}
\label{fig:generalspinorbits}
\end{figure*}

\begin{figure*}[!ht] % 2-column figure 
\centering
\vskip0.8cm
\includegraphics[width=8.3cm,viewport=84 588 320 780]{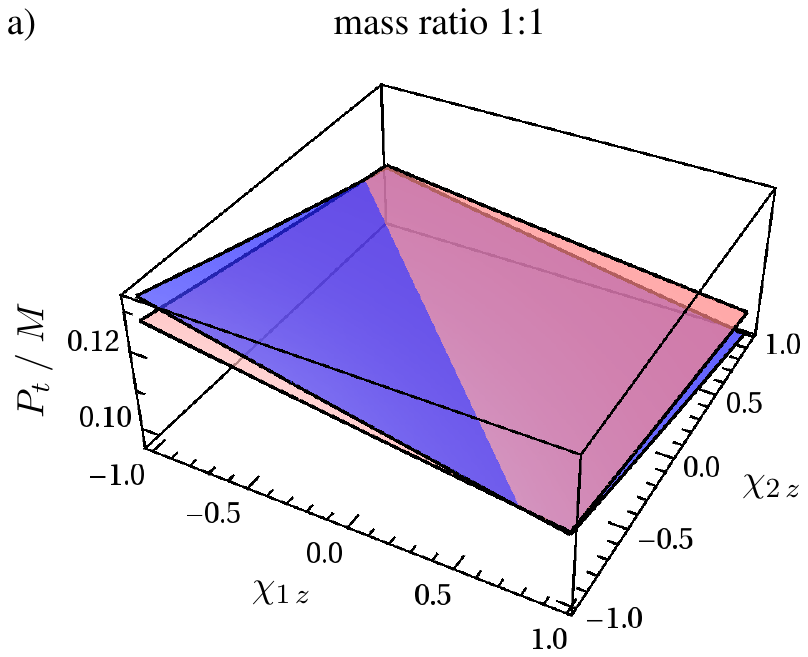}
\hskip0.8cm
\includegraphics[width=8.3cm,viewport=84 588 320 780]{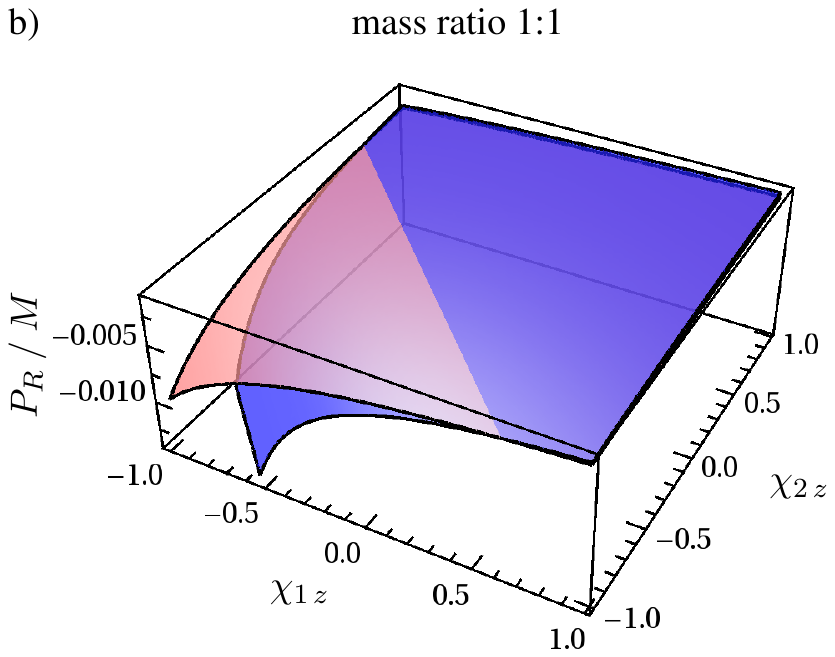}
\\[1.4cm]
\includegraphics[width=8.3cm,viewport=84 588 320 780]{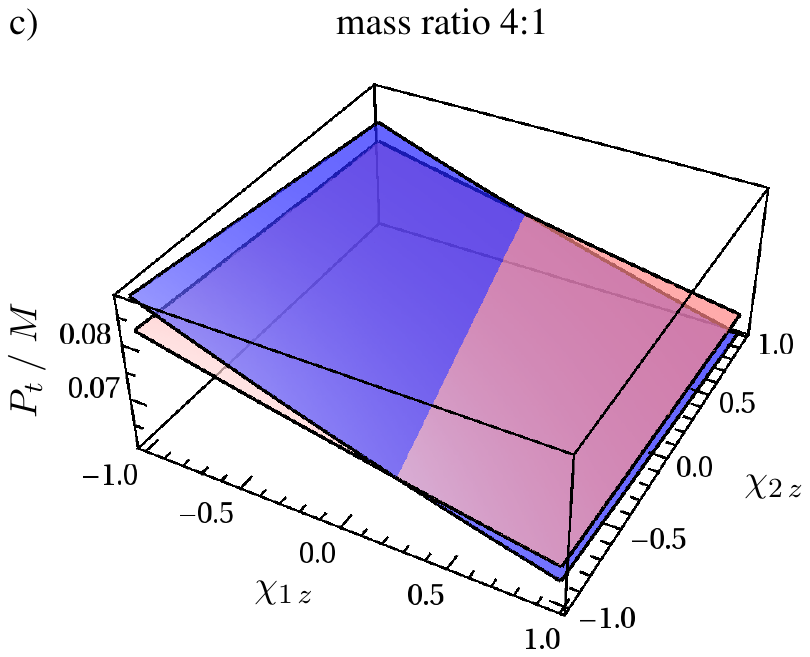}
\hskip0.8cm
\includegraphics[width=8.3cm,viewport=84 588 320 780]{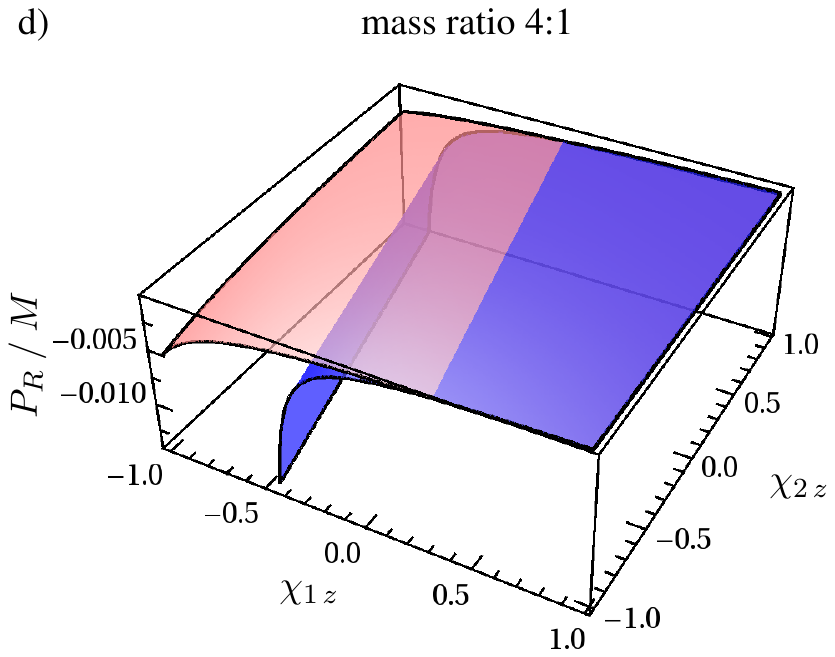}
\caption{
(Color online)
Modifications of PN initial parameters $P_t$ and $P_R$ for a binary at
a fixed initial separation of $8M$ when spins are involved. The plots
show THTF (blue/dark) and EHTF (red/light) initial data surfaces, for
which the $z$-component of the spins is varied in the interval
$[-1,+1]$. The zero-spin case marks the center of the plots, while
e.g.\ the $\chi_{1\,z}=\chi_{2\,z}=1$ case corresponds to
maximally-spinning black holes with spins aligned to the orbital
angular momentum. Changing to Pad\'{e}-resummed flux does not cause
visible modifications of the surfaces.}
\label{fig:spinparameter}
\end{figure*}

As a first step towards exploring more general spin configurations, we
consider one particular example with arbitrarily chosen spin vectors
in Fig.~\ref{fig:spineccentricity}\,f. In this case the motion does
not remain in the $z=0$ plane and the orbital plane shows considerable
precession, see Fig.~\ref{fig:generalspinorbits}. Both spins are of
magnitude $\chi=0.5$ with a smaller but positive component in the
$z$-direction, and the eccentricity is indeed in the range that we
would expect from the graphs Fig.~\ref{fig:spineccentricity}\,c-d
where spins of magnitude $0.25$ and $0.5$ are aligned to the
$z$-direction. This indicates that our initial data are not restricted
to aligned spins only, but also lead to small eccentricity for more
general spins.

Finally, let us look at the spin-spin interaction issue once more.  As
stated in the discussion of the aligned spin setup, the spin-spin
interaction term simplifies in this case and might be incorporated in
the initial data computation.  This was done in
Fig.~\ref{fig:spineccentricityenlarge} and indeed eccentricity is
reduced by a factor two for the EH data. This approach, however,
cannot amend the poor performance of the TH initial parameters.

After noting the effect of the different initial data types, let
us investigate how the large differences can be explained by having a
closer look at the particular values of $P_t$ and $P_R$. In
the equal-mass setup without spins the differences among the initial
parameters were comparatively small even up to close binary
separations, Figs.~\ref{fig:ptangcompareadmtoeob} and
\ref{fig:pradcompareadmtoeob}. When incorporating spins (parallel to
$\bs{L}$) we observe the modifications to tangential and radial
momentum shown in Fig.~\ref{fig:spinparameter}. 
As a first finding the tangential component experiences significant
changes all over the spin-parameter space. This can be explained by
means of frame dragging. Spacetime is twisted in the vicinity of each
spinning hole and the other object is forced to follow this
twist. When both spins are aligned with the orbital angular momentum
each black hole tries to accelerate the other on its orbit, with the
effect that less orbital momentum $P_t$ is needed for a
circular inspiral. (Recall that it is the velocity which is
responsible for the orbit's shape and that there is a complicated,
spin-dependent relation between 
$\dot{\bs{X}}=\partial H(\bs{X},\bs{P},\bs{S}_1,\bs{S}_2)/\partial\bs{P}$ 
and the momentum.) 

The TH and EH surfaces almost coincide in the
non-spinning case, which marks the center of the plots. With
increasing spins the Taylor-expanded data exhibit a stronger reaction
than the EOB ones. Recall that
comparatively small changes in $P_t$ can have strong effects on the
orbit. The TH and EH surfaces for $P_R$ are
almost identical for a certain part of the parameter space, in
particular as long as both spins are positive. 
For strong negative spins, especially for the dominant spin of the bigger
black hole at mass ratio 4:1, large changes in $P_R$ occur. Obviously,
the Taylor-expanded data are affected most, resulting in a behavior
similar to Fig.~\ref{fig:pradcompareadmtoeobmassratios}\,d, such that
no reasonable solution is obtained beyond a certain spin
threshold. Considering their poor performance in numerical runs, we
conclude that TH data are as inappropriate in this regime as in the
non-spinning case at high mass ratios.

The EOB-based data do surprisingly well recalling the
difficulties that arise from the transformation
EOB $\rightarrow$ ADM. The systematic simplifications we imposed, in
particular the approximation concerning the binary separation $R'$
entering the spin-orbit Hamiltonian, Eq.~(\ref{eq:hso})
resp.\ (\ref{eq:spinorbithamiltonian}), do not seem to significantly
degrade their performance.

%%%%%%%%%%%%%%%%%%%%%%%%%%%%%%%%%%%%%%%%%%%%%%%%%%%%%%%%%%%%%%%%%%%%%%%%%
\section{Conclusions}

Based on the work of \cite{BuoCheDam05} we presented several
variations of an (almost) analytical algorithm capable of delivering
quasi-spherical initial data for black-hole binary systems. By
employing two different PN Hamiltonians in combination with two
versions of energy-flux functions we constructed four types of initial
parameters. For the resulting initial data we performed numerical
simulations within the moving puncture framework with the BAM code,
both for non-spinning and spinning binaries, where we focussed on the
class of aligned spins. An eccentricity measurement provided
information on the quality of the data. In summary, the eccentricity
of the initial data considered grows for smaller binary
separations, larger mass ratios and larger spin magnitudes.

In the non-spinning case, differences in the initial parameters $P_t$
and $P_R$ as well as in the resulting eccentricities were found to be
comparatively small and often beyond the accuracy of our
measurement. However, TH data happen to suffer from a parasitic pole
at higher mass ratios. In particular due to the EH data we were able
to meet or even to obtain slightly lower eccentricities than those
resulting from the PN-evolution method of \cite{HusHanGon07}. 
As in \cite{HusHanGon07}, a non-zero radial component of the initial
momentum is required for best results.

As soon as spins are incorporated, PN theory shows that non-eccentric
inspirals are not possible in general due to spin-spin interactions,
but we demonstrated that EH initial data are suitable to minimize the
radial oscillations to a certain degree, at least for special,
aligned spin constellations. We considered one example for non-aligned
spins, for which a similarly small eccentricity was obtained.  Further
runs with arbitrary spin directions are planned for future
research. As expected, the eccentricities in simulations with spins
turned out to be larger than the spinless ones. The TH data gave
comparatively poor results, which again can be considered an effect of
the pole mentioned above.

To end with a concrete suggestion for practical implementations, a
good starting point is the EOB Hamiltonian with the Taylor expanded
flux, especially for non-zero spin, but also for mass ratios 4:1 and
larger. For smaller mass ratios the Taylor Hamiltonian gives smaller
eccentricities, but the eccentricity for both EH and TH is quite
small.
The main complication in the EOB method is the need for the
transformation to ADMTT coordinates for the numerical initial data.
Table \ref{tab:initialdata} gives explicit values for initial
parameters for several non-spinning binary configurations that can be
used directly in simulations.  Use the combination of $P_t$ and $P_R$
for which the eccentricity given in the last column of the table is
lowest.
%For example, for a run with mass ratio 4:1 and an initial separation of $10M$,
%use the data given in the ``EHPF'' row of table \ref{tab:initialdata}.
Initial data for several binary configurations with (anti-)aligned
spins can be read off table \ref{tab:initialdataconstraints}. For all
cases with $\chi \neq 0$, the EHPF data leads to the lowest
eccentricity.

Further investigations suggest themselves.
The PN-evolution method
of~\cite{HusHanGon07} depends on the quality and characteristics of
the Hamiltonian evolution method. It is natural to revisit this topic
with the insights gained in the present work for unequal masses and
spins, and also to compare with the PN-evolution methods
of~\cite{CamLouNak08,HusHan08}. 
Our study centers on the PN method of~\cite{BuoCheDam05} and numerical
evolutions. A refined method called ``post-post-circular'' initial
data has been suggested in~\cite{DamNagDor07}, which is worth investigating
in the context of numerical evolutions as well.

Apart from evolutions, it could be interesting to compare to other
diagnostics for the eccentricity of orbits, e.g.\
\cite{MorWil03,BerIyeWil06,GriCoo07}.

An alternative to the TH and EH methods considered here is the
semi-analytic puncture evolution approach to model inspiralling black
holes~\cite{GopSch08}. It is natural to expect that the resulting orbital
parameters are well adapted to numerical puncture evolutions, and we plan to
carry out a quantitative study.
%%%%%%%%%%%%%%%%%%%%%%%%%%%%%%%%%%%%%%%%%%%%%%%%%%%%%%%%%%%%%%%%%%%%%%%%%
\acknowledgments
It is a pleasure to thank Achamveedu Gopakumar, Mark Hannam, Sascha
Husa, and Gerhard Sch\"{a}fer for discussions and valuable insights in
the PN method.
This work was supported in part by DFG grant SFB/Transregio~7
``Gravitational Wave Astronomy'' and the DLR (Deutsches Zentrum f\"ur Luft
und Raumfahrt).
Computations were performed on the HLRB2 at LRZ Munich.
%#########################################################################

%\bibliography{refs,refsextra}
%%%%%%%%%%%%%%%%%%%%%%%%%%%%%%%%%%%%%%%%%%%%%%%%%%%%%%%%%%%%%%%%%%%%%%%%%

% all figures of the paper

\end{document}